\documentclass[a4paper,11pt]{article}
\pdfoutput=1 
\usepackage{jinstpub} 
\usepackage{multirow}
\usepackage[above, below]{placeins}

\title{Effects of misalignment on response uniformity of SiPM-on-tile technology for highly granular calorimeters}

 \author[1]{L.M.S.~de Silva\note{now at DESY, Hamburg, Germany.}}
 \author[2]{and F.~Simon\note{Corresponding author.}}
 \affiliation{Max-Planck-Institut f\"ur Physik,\\F\"ohringer Ring 6, 80805 M\"unchen, Germany}

\emailAdd{fsimon@mpp.mpg.de}

\abstract{The SiPM-on-tile technology, consisting of plastic scintillator tiles with typical sizes of a few cm$^2$ mounted on top of SiPMs, has been developed within the CALICE collaboration to enable automatized mass production of active elements for scintillator-based highly granular calorimeters. We present a study of the impact of misalignment of the scintillator tile with respect to the photon sensor on the response uniformity and on the absolute light yield for square, hexagonal and rhomboidal scintillator tiles of different sizes. A misalignment results in the formation of a dipole asymmetry of the spatial distribution of the light yield, with a magnitude that scales linearly with the size of the displacement, while the average light yield over the full active area of the tile is not affected. These results provide guidance for the definition of tolerances for the production and assembly of large calorimeter systems, showing that an alignment precision of approximately \mbox{500 $\upmu$m} or better allows to limit the impact to acceptable levels.}


\begin{document}
\maketitle
\flushbottom

\section{Introduction}
\label{sec:Intro}

The SiPM-on-tile technology enables large, highly granular scintillator-based calorimeter systems thanks to its inherent scalability and amenability for mass production. This technology has been originally developed by the CALICE collaboration as an evolution of the physics prototype of the Analog Hadron Calorimeter (AHCAL) \cite{Adloff:2010hb}, and has been successfully applied in the AHCAL technological prototype \cite{Sefkow:2018rhp}. The technology has also been adopted for the scintillator part of the CMS high granularity calorimeter (HGCAL) \cite{Collaboration:2293646} and is considered for other applications, such as for electromagnetic calorimetry for near detectors in long baseline neutrino experiments \cite{Emberger:2018pgr}.

The SiPM-on-tile technology is based on silicon photomultipliers directly mounted on electronics boards as SMD components via standard PCB assembly techniques, and on scintillator tiles optimized for direct coupling to the SiPM, mounted on top of the sensors on the electronics board \cite{Blazey:2009zz, Liu:2015cpe}. To achieve a uniform response over the full active area of the scintillator tiles, specific shapes of the scintillator in the form of dimple at the position of the photon sensor, which is coupled via an air gap, are required \cite{Blazey:2009zz, Simon:2010hf}.

A key advantage of direct coupling of the photon sensor to the scintillator tile compared to the use of a wavelength shifting fiber for light collection embedded inside the scintillator material are reduced requirements of the alignment of the photon sensor with respect to the scintillator element. Nevertheless, the light illumination within the dimple is not perfectly uniform, and also shows a dependence on the point of origin of the scintillation photons within the scintillator volume. Thus, the precision of the relative alignment between the scintillator tile and the photon sensor has an impact on the response, in particular on the response uniformity, and as such is relevant for the performance of SiPM-on-tile based detectors. Understanding the impact of a misalignment is important for the establishment of tolerance requirements for large-scale productions of calorimeter active elements based on this technology. 

In practice, the misalignment will have two main sources. The first, and dominating source is the precision of the placement of the scintillator tile on the electronics board hosting the photon sensors, which is connected to the precision of the machine used for this placement as well as to the accuracy of the dimensions of the tiles. Typical inaccuracies here are several 100 $\upmu$m. The second source is the accuracy of the position of the photon sensor, which is influenced by the precision of the component placement on the board as well as effects of the soldering procedure, where movement of the sensor during reflow soldering due to surface tension effects of the molten solder results in limits on the overall position accuracy. For the strictest tolerances for SMD assembly according to IPC Class 3, the specifications require a final accuracy in the order of 100 $\upmu$m for the typical pad dimensions used for the mounting of the photon sensors. Together, this defines the range of misalignment to be studied, suggesting a  region of approximately $\pm$1 mm, with step sizes in the order of 100 $\upmu$m.

\section{Study of scintillator tile misalignment}
\label{sec:Study}

To explore the impact of a misalignment of the coupling dimple with respect to the photon sensor, the response uniformity over the active area of different scintillator tiles for varying displacements has been studied. At this point, this is purely based on laboratory measurements. Simulation studies, which have been successfully used to model response non-uniformities depending on scintillator shape \cite{Liu:2015cpe}, have not been performed in the context of this work. The response is measured using a two-dimensional scanning setup with a radioactive source, as described in the following. 

\subsection{The experimental setup}
\label{ssec:Setup}

\begin{figure}
    \centering
    \includegraphics[width=0.48\textwidth]{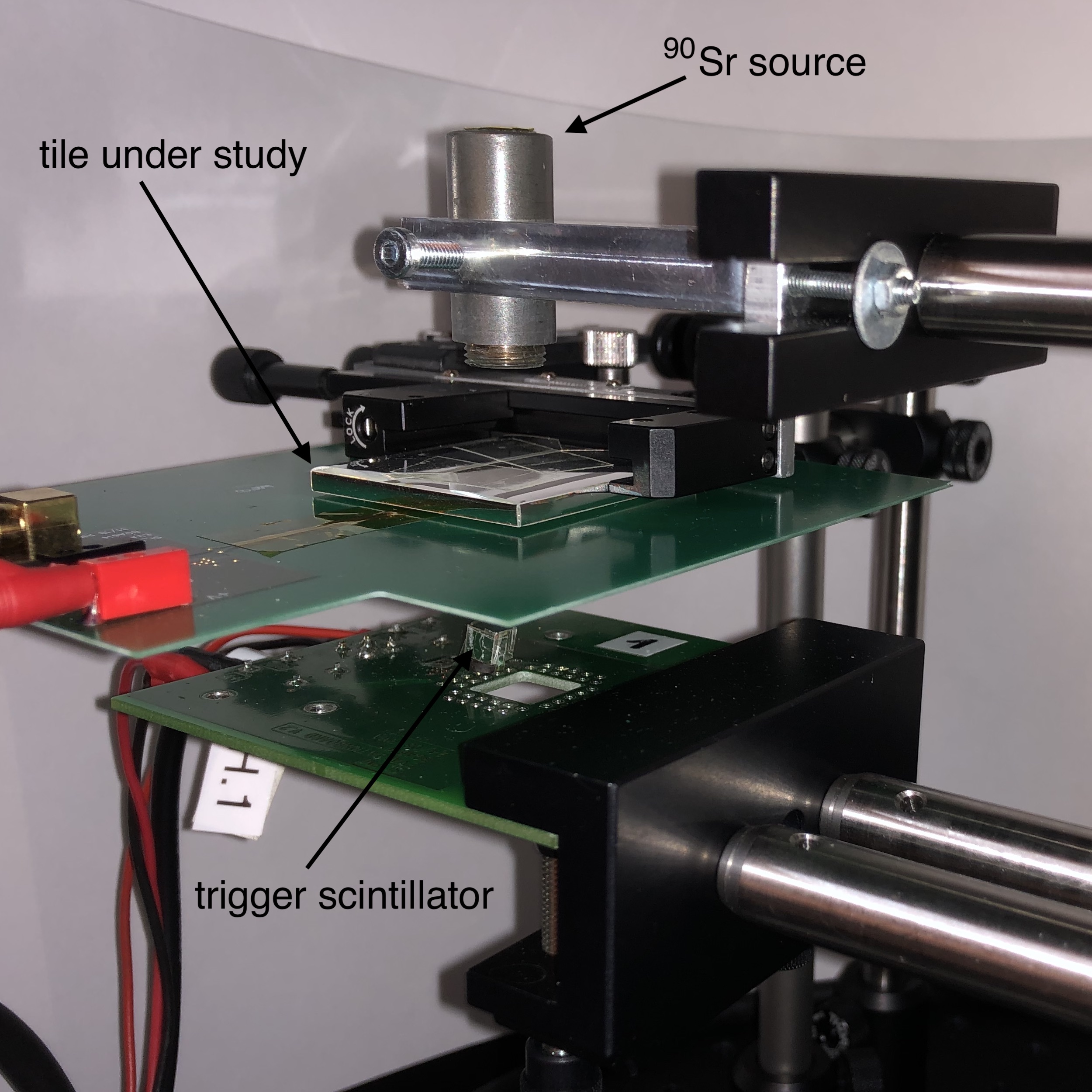}
    \caption{The experimental setup, showing the tile under study as well as the radioactive source above and the scintillator tile under the tile. The latter two are moved by precision translation stages during the scanning. Note that the white background behind the setup is introduced to improve visibility, during scanning operations the interior of the climate chamber is covered with black sheets to avoid noise from stray light.}
    \label{fig:Setup}
\end{figure}

All measurements are performed with a computer-controlled two-dimensional scanning setup that moves the tile under study in 1 or 2 mm steps between a $^{90}$Sr source above the tile and a $5 \times 5 \times 5$ mm$^3$ trigger scintillator below the tile. The main structure of the setup is described in \cite{Simon:2010hf}. It has been extended with additional components providing improved movement and full climate control \cite{Munwes:2634923}.  Figure \ref{fig:Setup} shows the experimental setup, installed in a climate chamber. 

\begin{figure}
    \centering
    \includegraphics[width=0.48\textwidth]{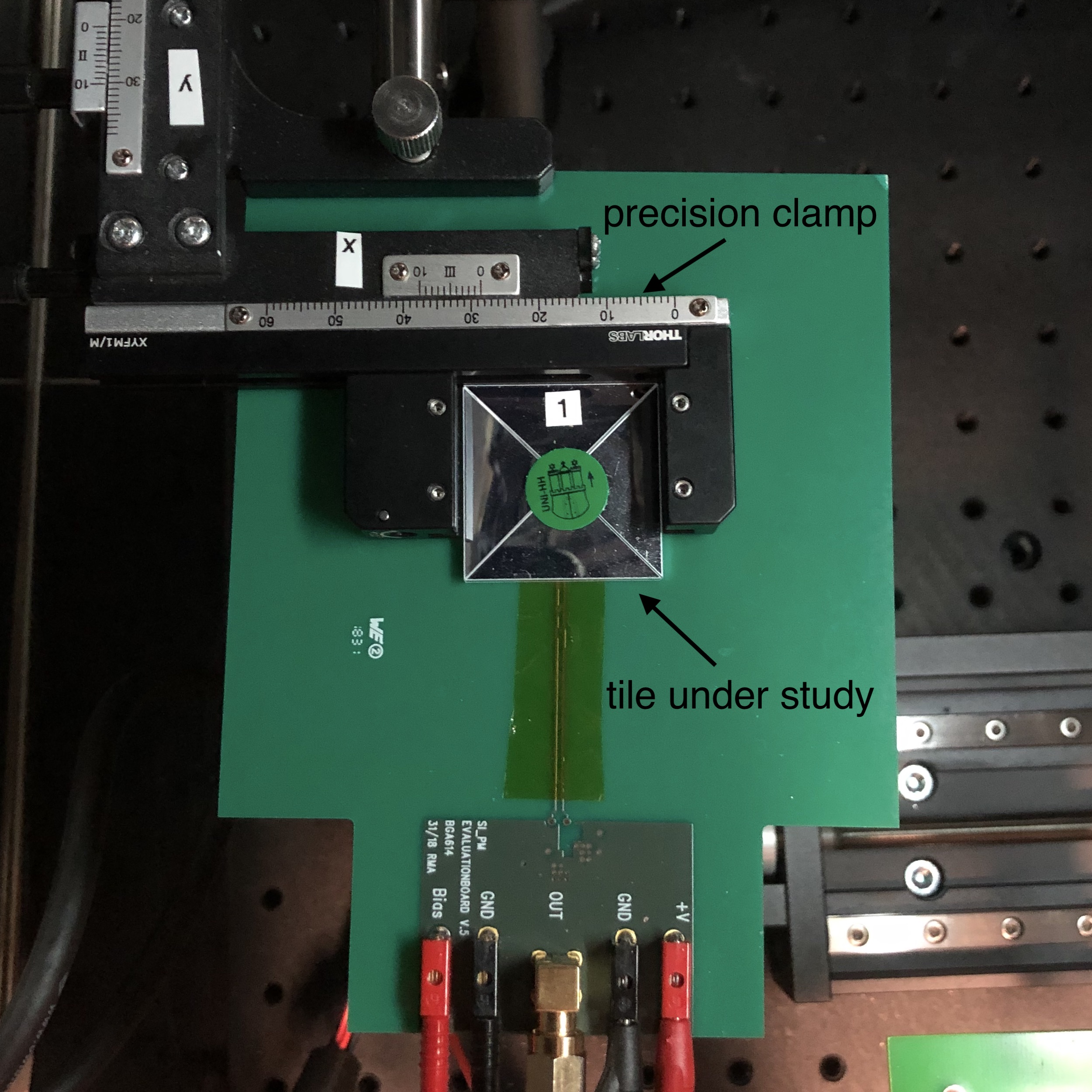}
    \hfill
    \includegraphics[width=0.48\textwidth]{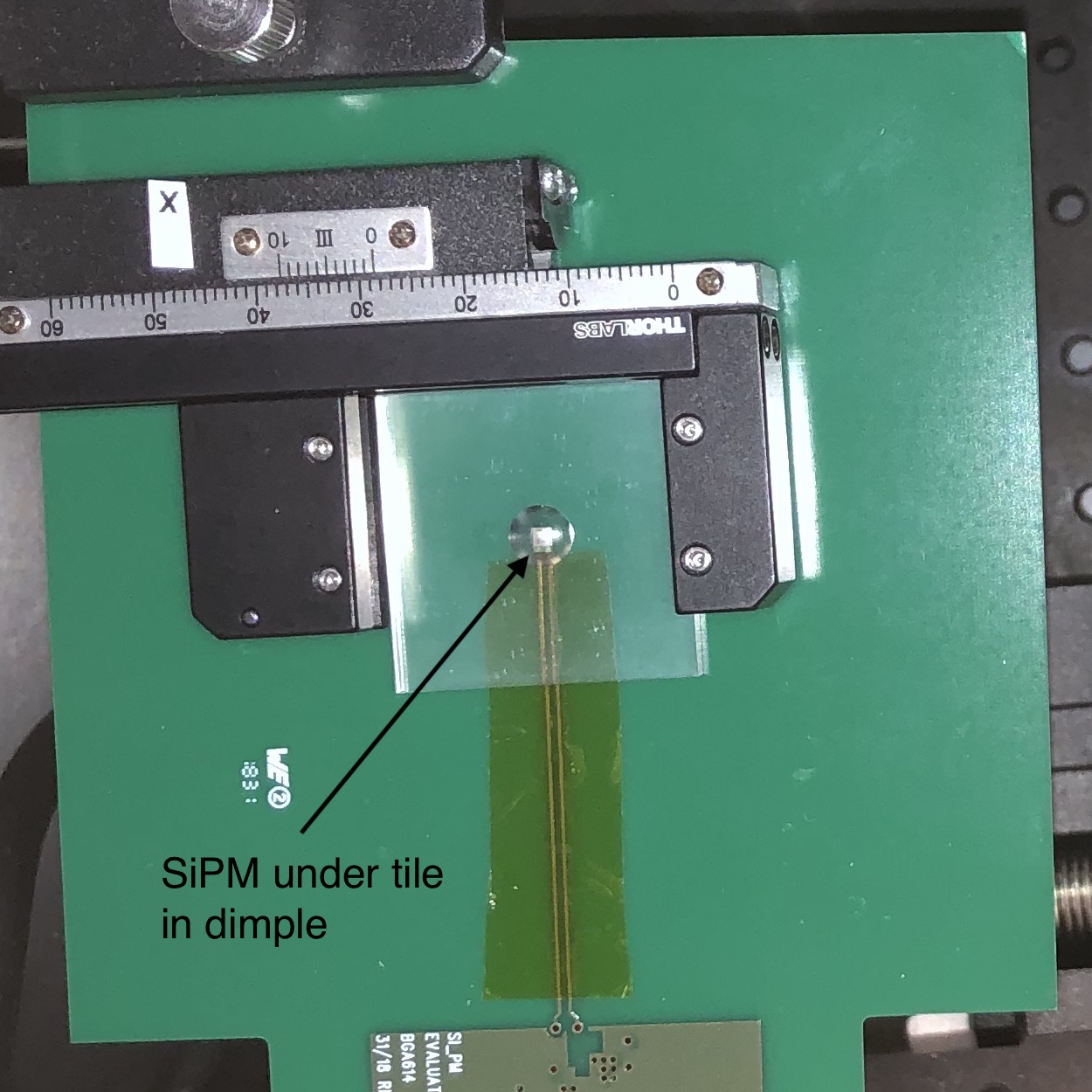}
    \caption{Details of the experimental setup, showing the adjustable precision clamp holding the tile under study (here the CALICE AHCAL tile) on the left, and the photon sensor shown through an un-wrapped BC-408 30 $\times$ 30 mm$^2$ tile on the right.}
    \label{fig:SetupDetails}
\end{figure}

The scintillator tile under study is mounted on a printed circuit board which hosts a Hamamatsu MPPC of type S13360-1325PE\footnote{Data sheet available at \href{https://www.hamamatsu.com/resources/pdf/ssd/s13360_series_kapd1052e.pdf}{https://www.hamamatsu.com/resources/pdf/ssd/s13360\_series\_kapd1052e.pdf}}, with an active area of $1.3\,\times\,1.3$ mm$^2$ hosting 2668 pixels, as well as a preamplifier that drives the photon sensor signal over a coaxial cable to a computer-controlled digital oscilloscope. The scintillator tiles are held in place with an adjustable precision optical clamp mounted on the circuit board, allowing the manual adjustment of the tile position with approximately 50 $\upmu$m precision. These details are shown in figure \ref{fig:SetupDetails}.

The measurements discussed here are all performed at an operating voltage of 55 V, corresponding to an overvoltage of 3 V, at a temperature stabilized at 20 $^\circ$C. Under these conditions, the photon sensor has a photon detection efficiency (PDE) of $\sim$20\% for the relevant wavelength around 430 nm. At an overvoltage of 5 V, which is often used for these types of sensors, the PDE is approximately 30\% higher. This needs to be considered when comparing absolute light yield measurements presented here and elsewhere.

The effect of the misalignment is studied by displacing the tile along one axis with respect to the nominal position, where the apex of the dimple is centered on the active area of photon sensor. For each position, a full scan of the scintillator tile is performed, which takes approximately 8 hours. The scanning itself is performed fully automatically without external interventions, but the positioning of the tile with respect to the SiPM prior to each full scan is done manually. The long scanning time and the need for manual interventions limits the overall throughput to small sample sizes. To determine the optimal position of the tile with respect to the photon sensor, exploratory measurements prior to the systematic displacement of the tile are performed. The optimal position is taken as the one with a minimal observed asymmetry as described in section \ref{ssec:Misalignment}.

\subsection{Studied scintillator tile geometries}
\label{ssec:Tiles}

\begin{figure}
    \centering
    \includegraphics[width = 0.99\textwidth]{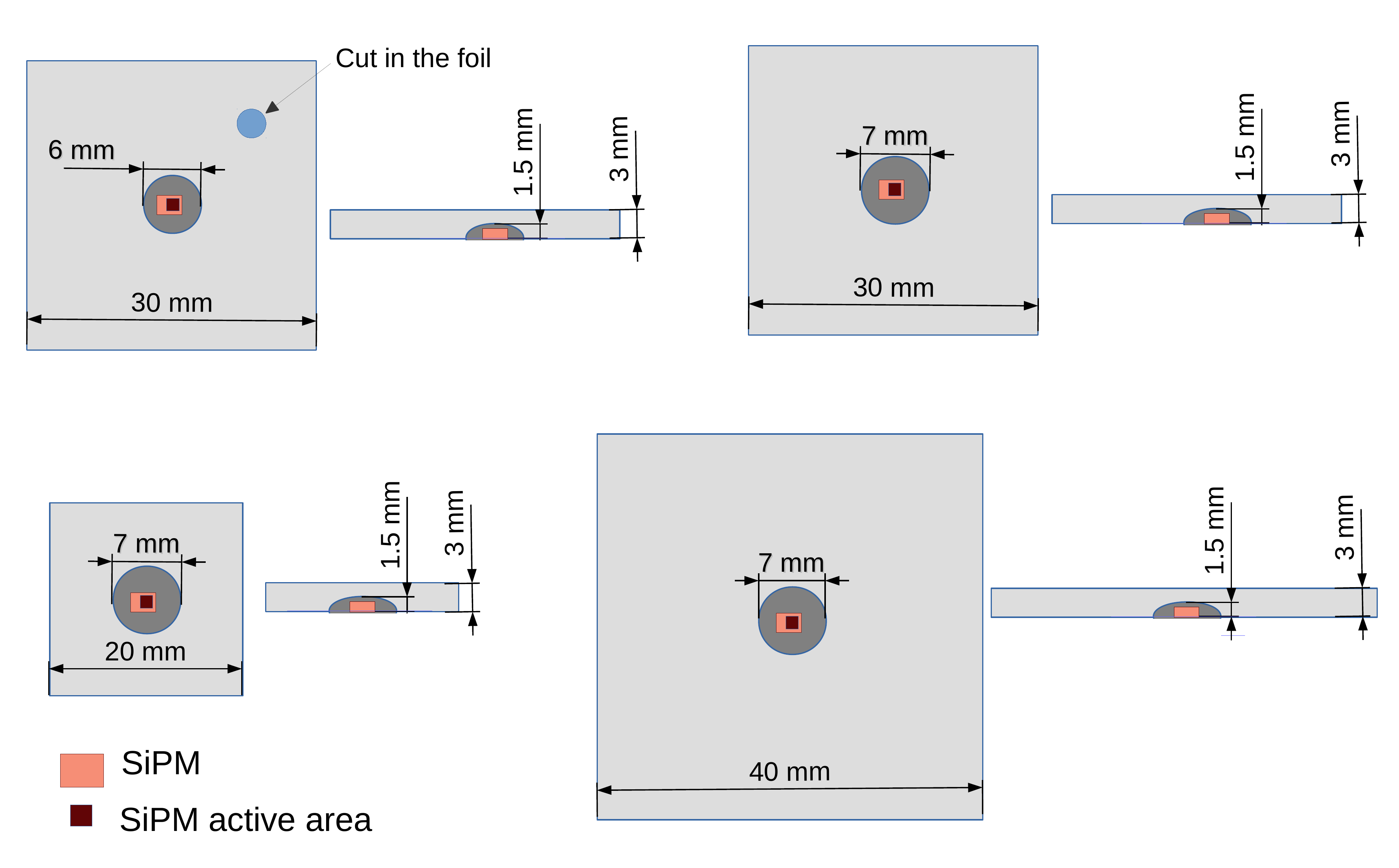}
    \caption{Illustration of the geometries of the square scintillator tiles studied. The nominal positioning of the SiPM inside of the dimple is also indicated, showing both the overall package dimensions and the active area of $1.3\,\times\,1.3$ mm$^2$. Top left: Injection-molded polystyrene-based scintillator tile used in the CALICE AHCAL technological prototype. Note the additional cutout in the reflective wrapping of the scintillator tile to admit light from a calibration LED. Top right: Machined square BC-408 scintillator tile with dimensions 30\,$\times$\,30\,mm$^2$. Bottom: Machined square BC-408 tile with dimensions 20\,$\times$\,20\,mm$^2$  (left) and 40\,$\times$\,40\,mm$^2$ (right).}
    \label{fig:TileGeometries}
\end{figure}

For the present study, square, hexagonal and rhomboidal scintillator tile geometries were studied. While the current projects including the CALICE AHCAL and the CMS HGCAL use square or near-square geometries, hexagonal and rhomboidal tiles offer potentially advantageous properties for endcap geometries, allowing the tiling of circular surfaces with uniform tiles over the full area rather than introducing the need for geometry variations as a function of radius. Hexagonal and rhomboidal tiles also are in general well adjusted to detector designs with hexagonal electronics boards as central elements defining the geometry.

The scintillator tiles used in the present study include the polystyrene-based injection-molded scintillator\footnote{Polystyrene base with 2\% p-terphenyl and 0.01\% POPOP, produced by Uniplast, Vladimir, Russia} tile used in the CALICE AHCAL technological prototype with a size of \mbox{30 $\times$ 30 mm$^2$} \cite{Chadeeva:2018ezg}, square tiles machined from Bicron BC-408 scintillator\footnote{Data sheet available at \href{https://www.crystals.saint-gobain.com/sites/imdf.crystals.com/files/documents/bc400-404-408-412-416-data-sheet.pdf}{https://www.crystals.saint-gobain.com/sites/imdf.crystals.com/files/documents/bc400-404-408-412-416-data-sheet.pdf}} with sizes of  20 $\times$ 20 mm$^2$,  \mbox{30 $\times$ 30 mm$^2$} and 40 $\times$ 40 mm$^2$, and two different hexagonal scintillator tiles and one rhomboidal tile machined from BC-408. 
Since the polystyrene-based tiles were taken from the mass production for the CALICE AHCAL technological prototype, only one size was available, as defined by the production mold. Only a single rhombus-shaped tile was machined as a pilot study.
All scintillator tiles have a thickness of 3 mm, and are wrapped in 3M Vikuiti\texttrademark\ ESR\footnote{Product information available at \href{https://multimedia.3m.com/mws/media/374730O/vikuiti-tm-esr-sales-literature.pdf}{https://multimedia.3m.com/mws/media/374730O/vikuiti-tm-esr-sales-literature.pdf}} reflector film. Figure \ref{fig:TileGeometries} illustrates the different geometries used for the square tiles, and gives the dimensions of the various elements. It also illustrates the nominal positioning of the SiPM with the center of the active area aligned with the center of the dimple, as well as the size of the overall package and of the active area. The dimple is a perfect sphere with a depth of 1.5 mm in the scintillator material. The dimple in the square BC-408 tiles for all three sizes was measured to be slightly wider than for the other tiles, which is attributed to the inherent uncertainty and technician-dependent variations in the manual machining and polishing process. The hexagonal and rhomboidal tiles are discussed separately below. 

\subsection{Quantification of misalignment effects}
\label{ssec:Misalignment}

\begin{figure}
    \centering
    \includegraphics[width = 0.475\textwidth]{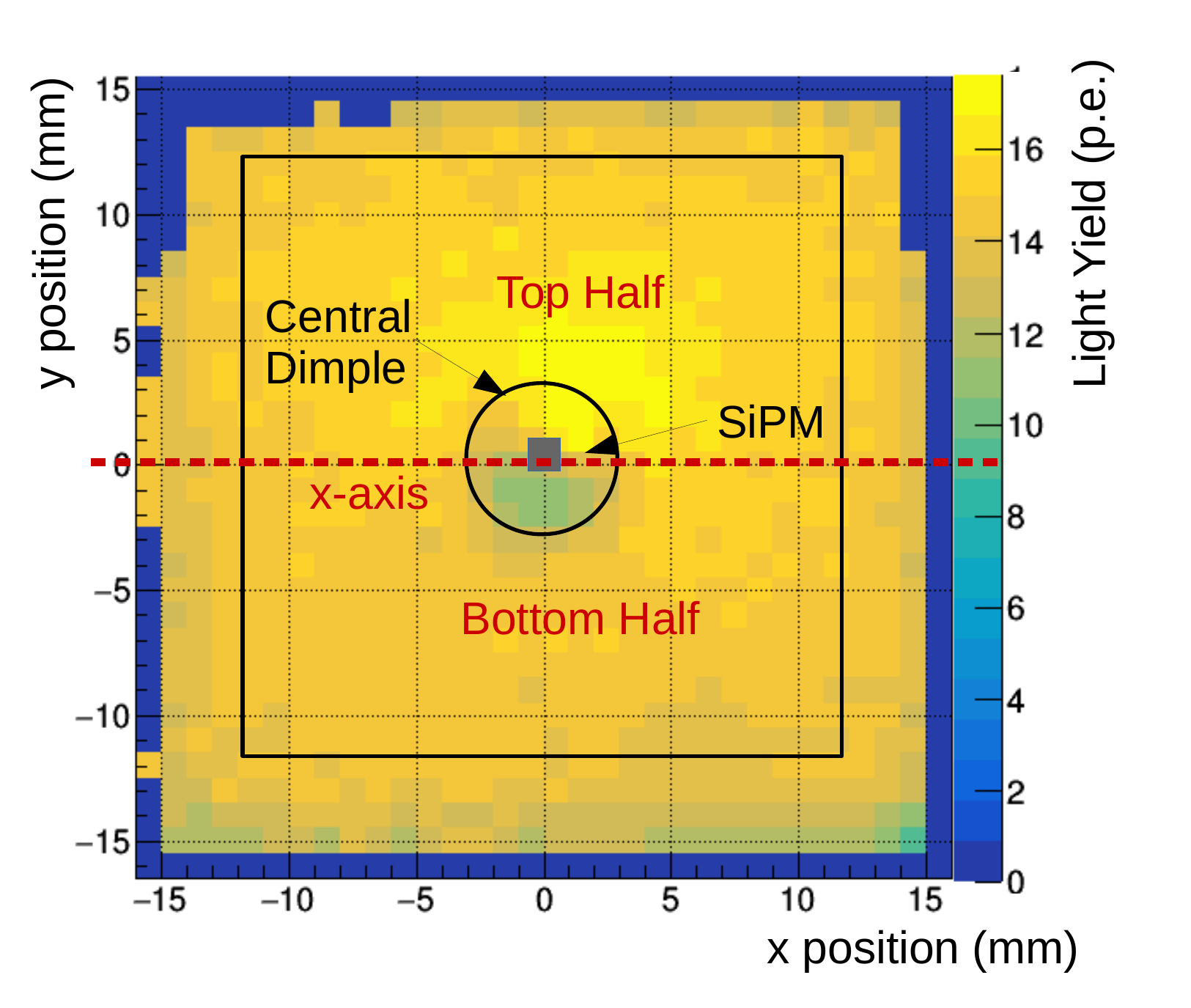}
    \includegraphics[width = 0.45\textwidth]{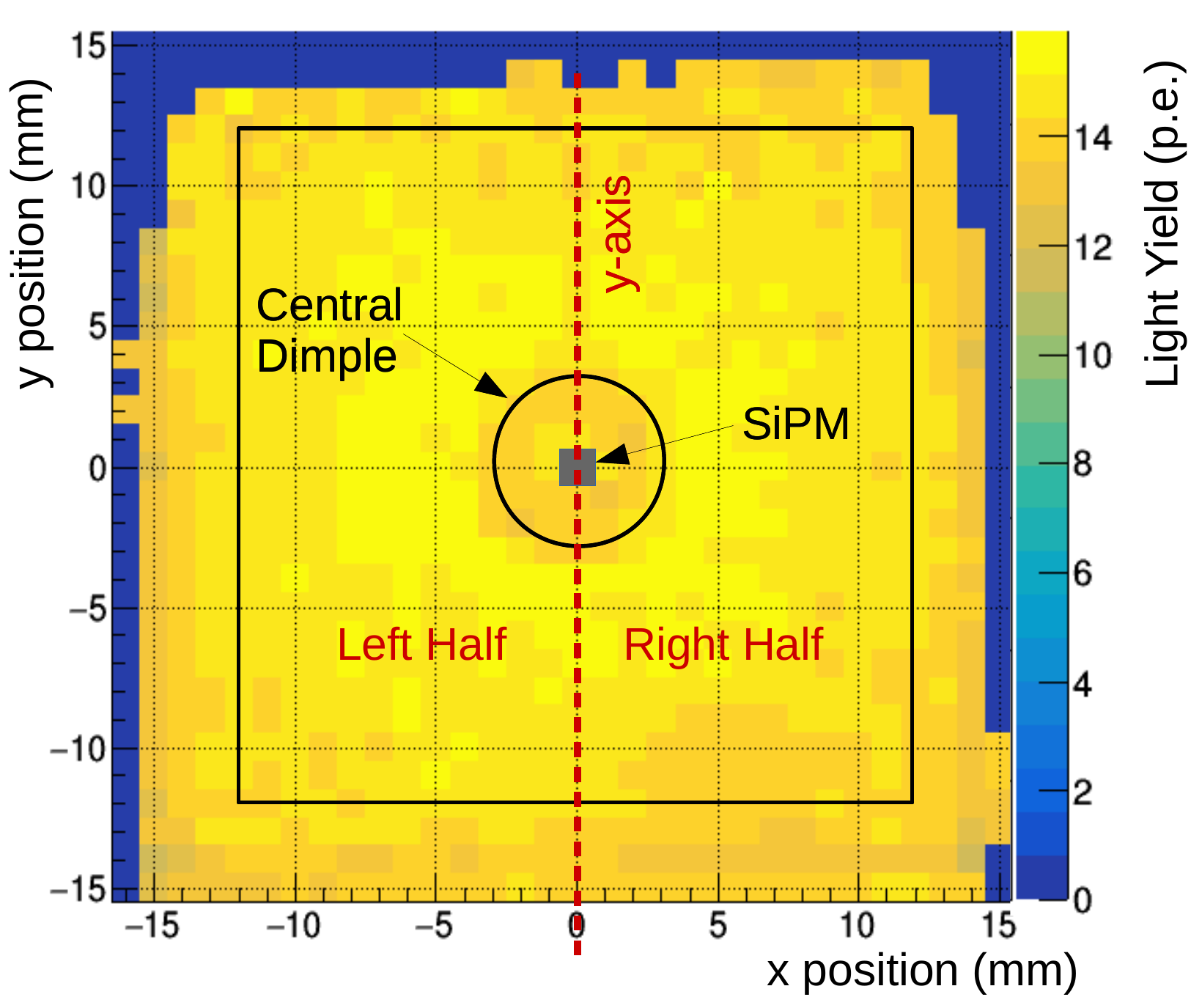}
    \caption{Scan results for the CALICE AHCAL scintillator tile, for a misalignment of approximately 1 mm (left) and for the perfect alignment with the photon sensor (right). The placement of the active area of the photon sensor (solid grey square), the region used to calculate the average response (black open square) and the axis used to define the two hemispheres for the calculation of the dipole asymmetry (dashed red line) is indicated on the figure. The response color scale is normalized to the bin with the highest signal.}
    \label{fig:CALICETileAlignment}
\end{figure}

The detailed scans of the response over the active area of the scintillator tiles have shown that the primary effect of a misalignment of the tile with respect to the photon sensor along one of the main axes of the tile is the formation of a dipole asymmetry in the response, as shown for the CALICE AHCAL tile in figure \ref{fig:CALICETileAlignment} left. Figure \ref{fig:CALICETileAlignment} right shows the case for perfect alignment. The color scale, which indicates the signal amplitude, is normalized to the highest bin. Note that the measurements in the edge regions are affected by the massive clamp, which fully absorbs electrons that impinge on it. Measurements at positions partially or fully under the clamp thus see a dramatically reduced trigger rate, and are distorted by otherwise rare events of scattered electrons and photons. This extends the apparent sensitive area of the scintillator tile at the left, right and bottom edge, as visible in figure \ref{fig:CALICETileAlignment}.  

To quantify the effect of a misalignment, a fiducial area is defined within the tile surface as indicated by the black square in the figure, centered on the nominal position of the photon sensor defined by the center of the dimple. Perpendicular to the direction of the displacement of the tile center with respect to the center of the photon sensor, the fiducial area is divided into a top and a bottom hemisphere, as indicated by the red dashed line in the figure. Using this subdivision of the active area of the tile, two quantities are calculated:
\begin{itemize}
    \item The average amplitude $S_{\mathrm{avg}}$, given by mean signal observed over the fiducial area.
    \item The asymmetry $A_{\mathrm{hem}}$, given by the difference of the two hemispheres, normalized to the average amplitude. This asymmetry is given by $A_{\mathrm{hem}} = (S_{\mathrm{avg, hem\, 1}} - S_{\mathrm{avg, hem\, 2}}) / S_{\mathrm{avg}}$, where $S_{\mathrm{avg, hem\, 1}}$ and $S_{\mathrm{avg, hem\, 2}}$ are the average amplitudes calculated over each of the two hemispheres. 
\end{itemize}

The asymmetry $A_{\mathrm{hem}}$ is taken as a measure for the effect of the misalignment of the tile with respect to the photon sensor. For a perfect alignment, the response pattern is expected to be symmetric, resulting in a vanishing asymmetry.

\section{Results}

The presentation of the results is separated by tile geometry, followed by a global discussion of observed trends and general features across all studied geometries.

\subsection{Square scintillator tiles}
\label{sec:SquareTiles}

For each of the square tile geometries presented in section \ref{ssec:Tiles}, two different scintillator tiles have been studied to provide a limited degree of control over variations from sample to sample. The results of the two measurements for each tile type are averaged to provide type-specific results. 

\begin{figure}
    \centering
    \includegraphics[width = 0.495\textwidth]{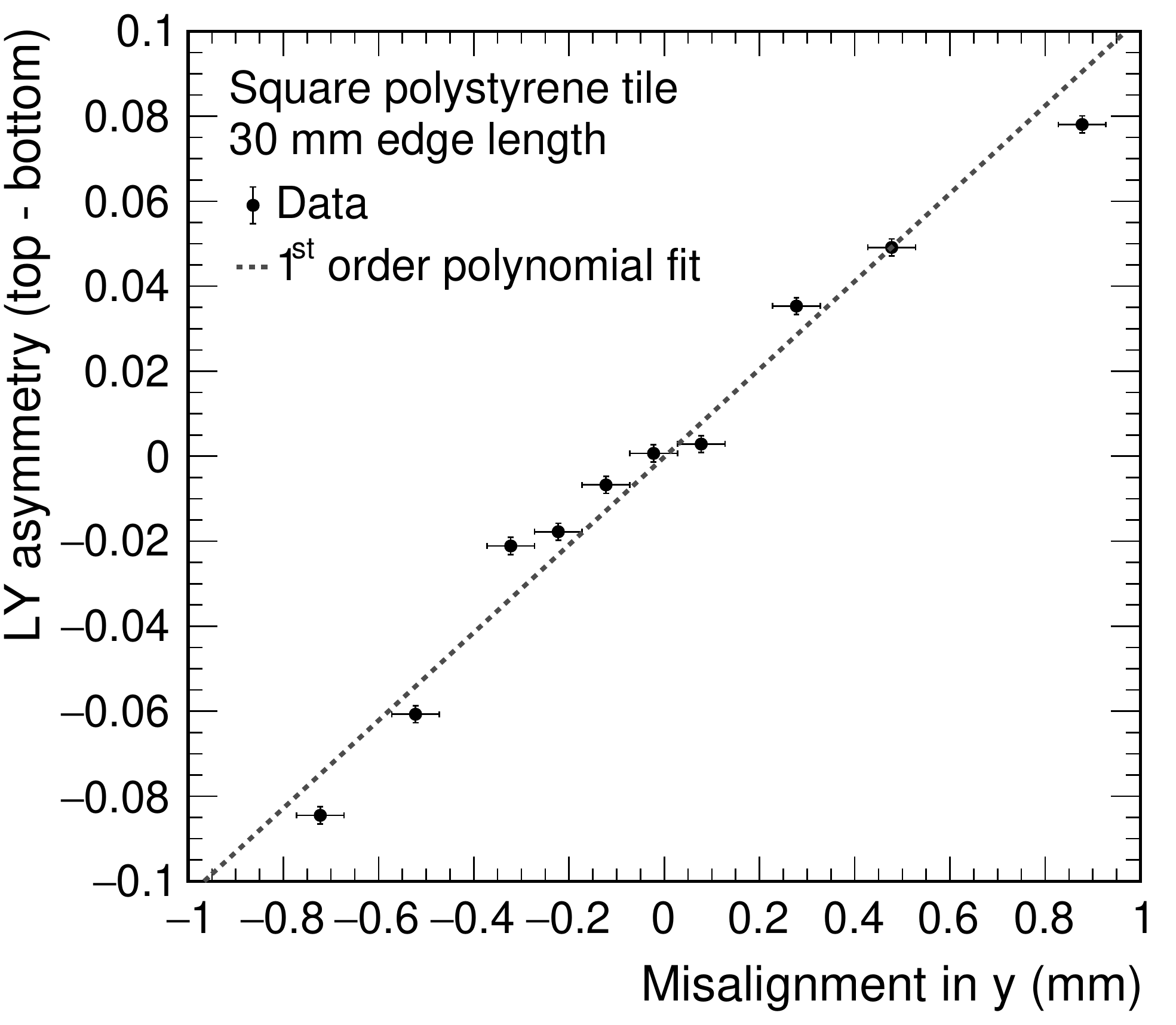}
    \includegraphics[width = 0.495\textwidth]{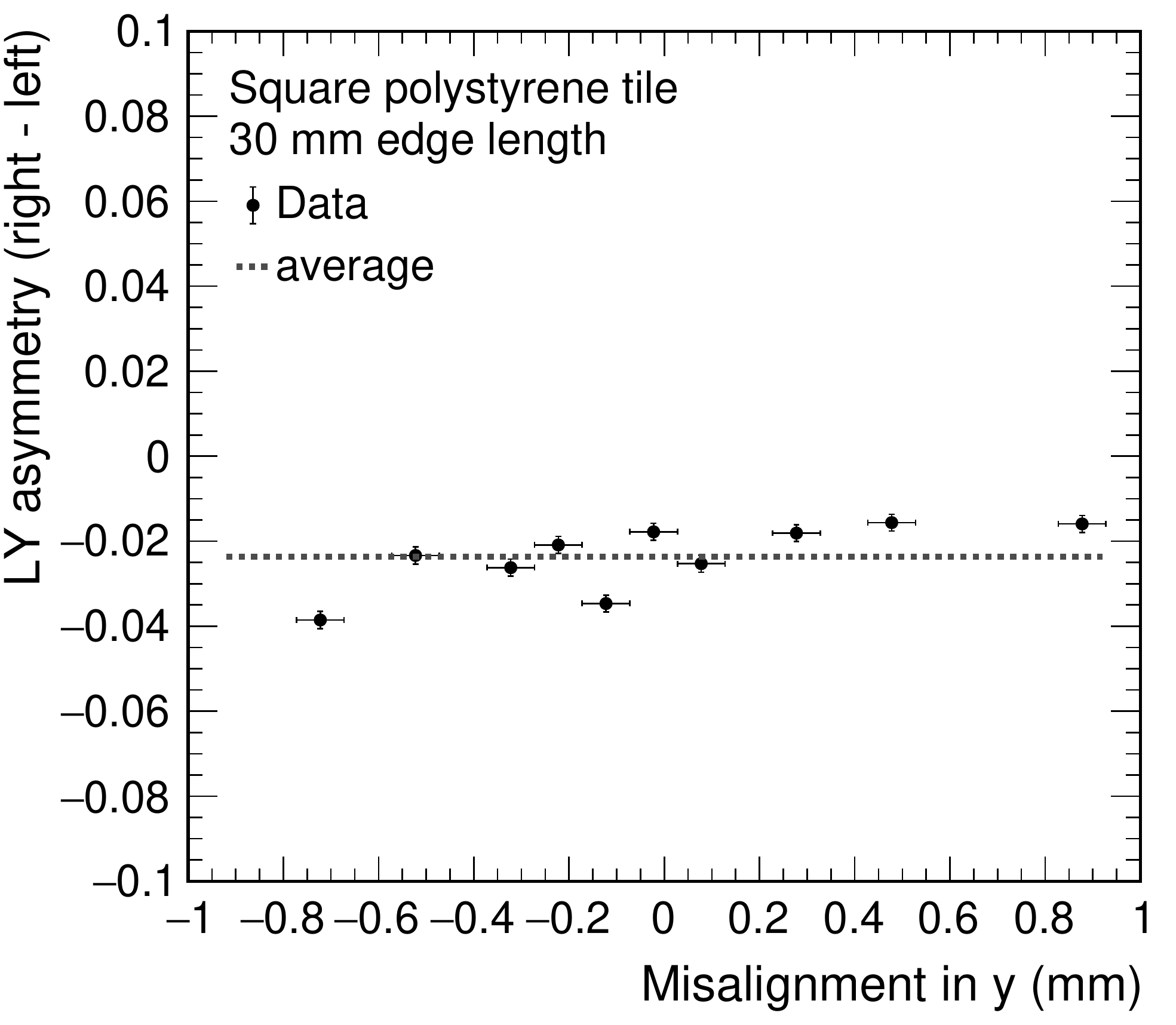}
    \caption{Measured light yield asymmetry $A_{\mathrm{hem}}$ for the polystyrene-based CALICE AHCAL tile for varying displacements in the $y$ direction. {\it Left:} Top - bottom asymmetry. The data points are fit with a first order polynomial to extract the slope as a measure for the impact of the displacement. The fitted slope is \mbox{$G_{norm}$  = 0.103 $\pm$ 0.006 mm$^{-1}$} . {\it Right:} Right - left asymmetry. The average of the data points is indicated by the horizontal line.}
    \label{fig:MeasuredAsymmetry}
\end{figure}

\begin{table}[htbp]
\centering
\caption{\label{tab:Results} Summary of the results of square tiles. For each tile geometry, two different tiles are measured, with the main result given by the average of the two measurements. The individual measurements are also reported. The uncertainties are the uncertainties of the fit. The PS tiles are wrapped by a robot, resulting in different characteristics, see text for details.}
\bigskip
\begin{tabular}{|l|c|c|}
\hline
Tile & average $G_{norm}$  [mm$^{-1}$] & $G_{norm}$  [mm$^{-1}$] \\
\hline
\multirow{2}{*}{BC-408 $20\times 20$ mm$^2$} & \multirow{2}{*}{0.127 $\pm$ 0.005} & 0.123 $\pm$ 0.005\\
& & 0.130 $\pm$ 0.009\\
\hline
\multirow{2}{*}{BC-408  $30\times 30$ mm$^2$} & \multirow{2}{*}{0.089 $\pm$ 0.004} & 0.091 $\pm$ 0.004\\
& & 0.087 $\pm$ 0.006\\
\hline
\multirow{2}{*}{BC-408  $40\times 40$ mm$^2$} & \multirow{2}{*}{0.066 $\pm$ 0.004} & 0.070 $\pm$ 0.004\\
& & 0.062 $\pm$ 0.006\\
\hline
\multirow{2}{*}{PS  $30\times 30$ mm$^2$} & \multirow{2}{*}{0.111 $\pm$ 0.004} & 0.103 $\pm$ 0.004\\
&  & 0.118 $\pm$ 0.007\\
\hline
\end{tabular}
\end{table}

Figure \ref{fig:MeasuredAsymmetry} shows one of the results of a series of measurements with different displacements of the center of the tile with respect to the photon sensor along one axis, performed for the polystyrene-based square CALICE AHCAL tile with a size of $30 \times 30$ mm$^2$. This measurement, as all other measurements performed, shows that there is a linear dependence of the observed asymmetry on the misalignment, when considering the asymmetry in the direction of the displacement. The response asymmetry in the direction orthogonal to the displacement is constant as a function of the displacement. For most measurements performed here, this asymmetry is consistent with zero. For the case shown in figure \ref{fig:MeasuredAsymmetry} {\it right} there is a small, constant non-zero asymmetry, with slightly higher light yield observed in the left half of the scintillator tile. This right-left asymmetry is due to a small displacement also in the $x$ direction on the order of 200 $\upmu$m, which originates from an imperfect adjustment of the central position of the tile along that axis prior to the displacement scan along the $y$ direction.

The impact of a misalignment on the cell response uniformity is quantified by the slope of the asymmetry with respect to the tile displacement, obtained by the fit of a first order polynomial to the data points, as  shown in figure \ref{fig:MeasuredAsymmetry} {\it left}. This fitted gradient of the normalized asymmetry per unit length of displacement is subsequently referred to as $G_{norm}$. A lower value for this slope corresponds to a smaller change in asymmetry per unit of displacement, indicating reduced susceptibility of a particular tile to misalignment. 
 Studies of displacements along both main axes of the scintillator tiles have shown that the effect of the misalignment is equal in size in both directions for square tiles, as expected.

The results for all studied square tiles are summarized in Table \ref{tab:Results}, giving the averaged non-uniformity slope as well as the individual measurements. The uncertainties are obtained from the fit, and include statistical uncertainties from the measurement of the signal amplitudes and uncertainties from the manual adjustment of the tile position assumed to be 50\,$\upmu$m. 

When comparing the BC-408 and the Polystyrene-based (PS) scintillator tiles, it is important to note that the PS tiles use a different method of wrapping. The PS tiles are taken from the mass production carried out for the construction of the CALICE AHCAL prototype \cite{Sefkow:2018rhp}, where they were wrapped with a robotic procedure in pre-cut 3M Vikuiti\texttrademark\ ESR reflective foil that closed on the top of the tile without overlap, potentially resulting in small gaps that may alter the light collection characteristics, and thus lead to changes in the impact of misalignment. To explore this, one of the two PS tiles was re-wrapped using the same foil design with manual wrapping as used for the BC-408 tiles. The slope of the asymmetry $G_{norm}$ of that tile changed from 0.118 $\pm$ 0.007 to 0.065 $\pm$ 0.003 with the manual wrapping. This observation underlines the important effect of the wrapping on the properties of directly coupled scintillator tiles. The fact that the slope for the PS tiles with manual wrapping is lower than the one observed for the equivalent BC-408 - based tiles indicates a material dependence of the impact of misalignment, likely connected to optical properties of the scintillator. A more systematic investigation of these effects is beyond the scope of the present paper.

\FloatBarrier

\subsection{Hexagonal and rhomboidal scintillator tiles}

\begin{figure}
    \centering
    \includegraphics[width = 0.495\textwidth]{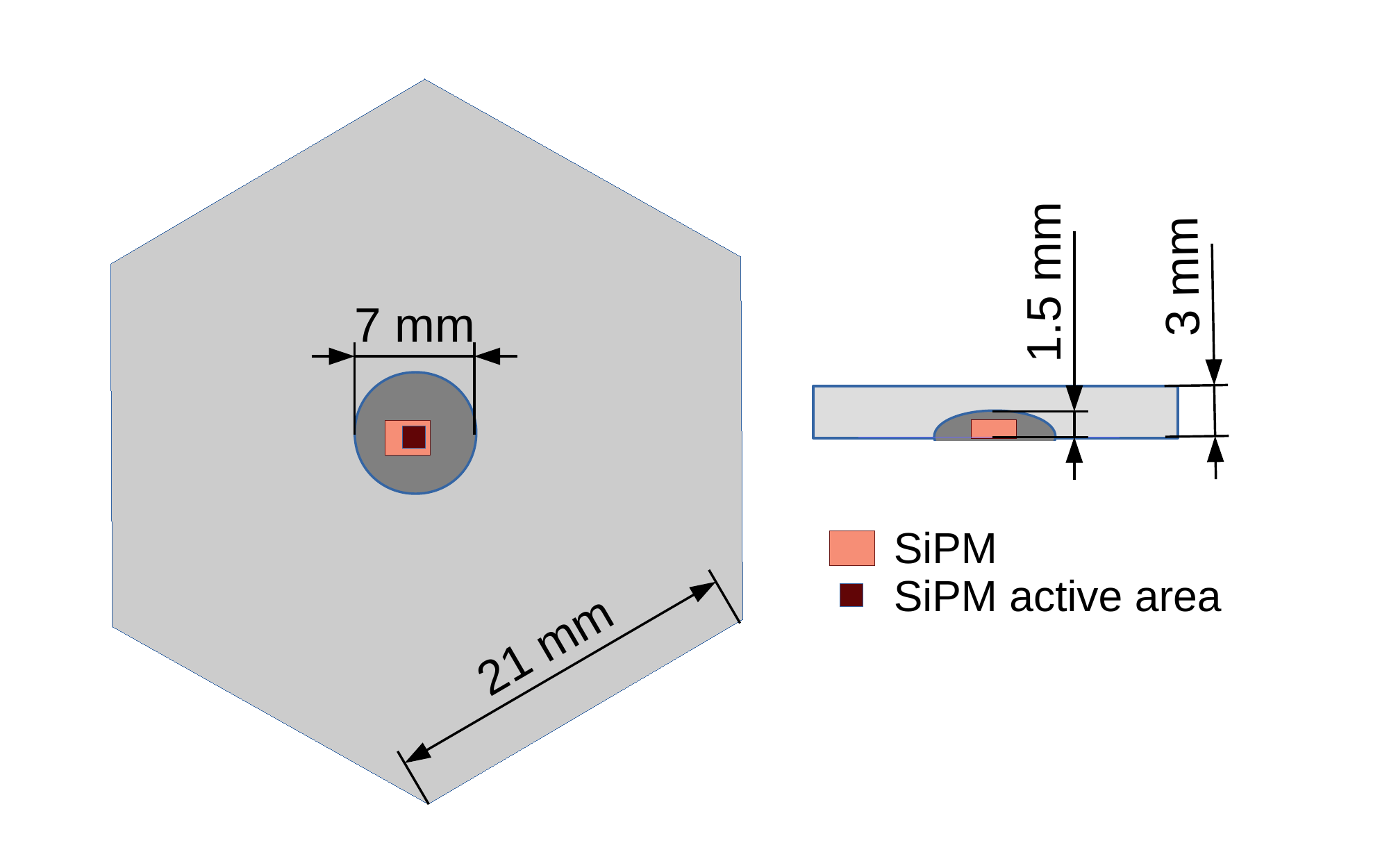}
    \hfill
    \includegraphics[width = 0.475\textwidth]{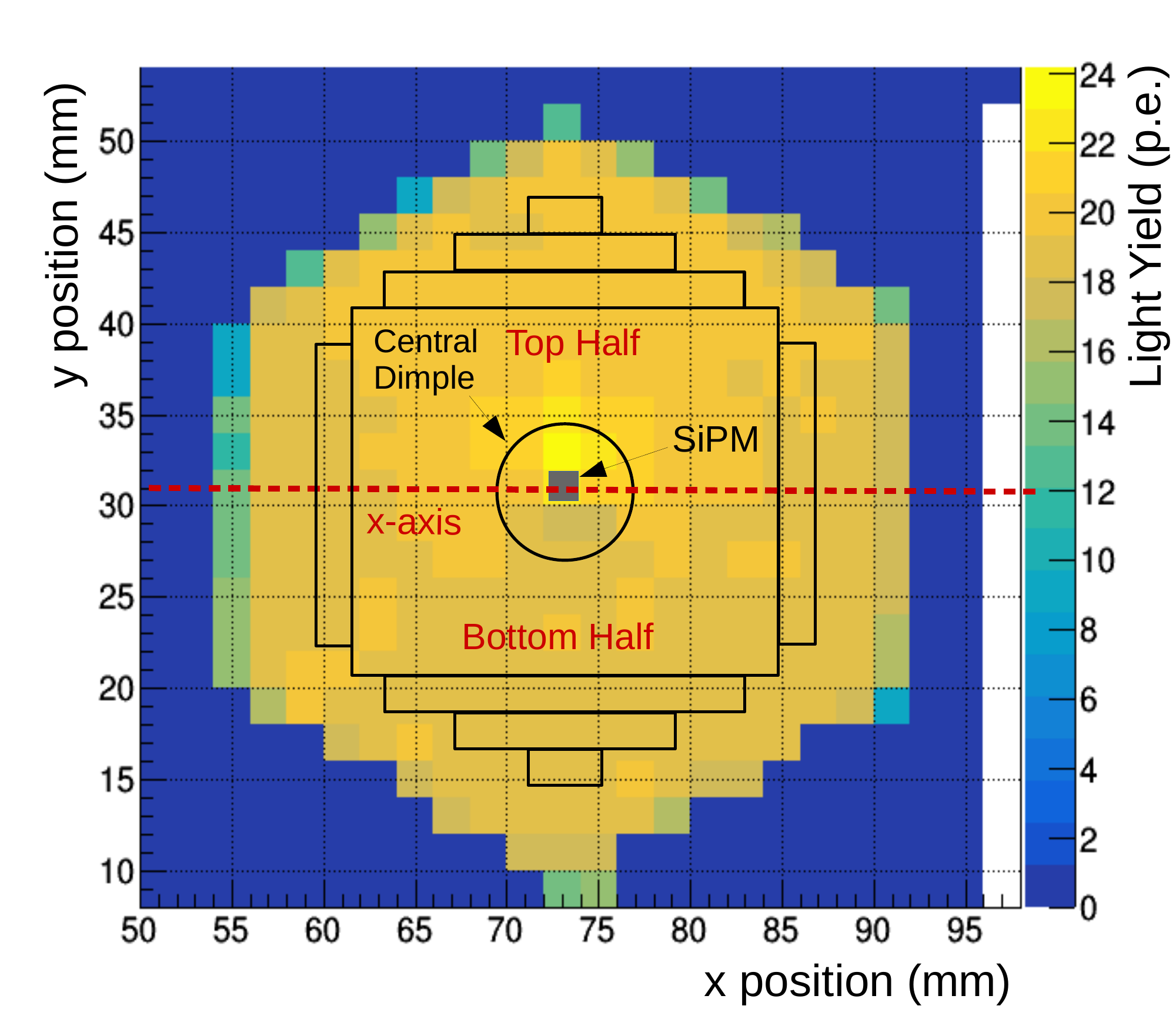}
    \caption{Illustration of the geometry including nominal SiPM placement (left) and scan results (right) for the hexagonal tile with an edge length of 21 mm. Also the rectangles used to define the fiducial area for the calculation of average response and asymmetry are shown. The response color scale is normalized to the bin with the highest signal.}
    \label{fig:HexTileAlignment}
\end{figure}

In addition to the square tiles, two different hexagonal tiles and one rhomboidal tile were studied using the same procedure as presented in detail above.

Figure \ref{fig:HexTileAlignment} shows the geometry of one of the two hexagonal tiles, together with a scan of the tile response. In addition, a smaller tile with an edge length of 14 mm instead of 21 mm was studied. All other properties of this tile are identical to the larger one shown in the figure. 
The figure also shows a marked difference between square and hexagonal tiles in the context of the displacements explored here. While the situation along the two main axes, horizontal and vertical in figure \ref{fig:CALICETileAlignment}, is identical for square tiles, for hexagonal tiles this is not the case, where one of the axes runs from the lower to the upper corner, while the other is perpendicular to the left and right edges. To accommodate this, the fiducial area used for the calculation of the average response and for the definition of the hemispheres is made up by the sum of a number of different rectangles, as illustrated in figure \ref{fig:HexTileAlignment}.  The definitions of the asymmetry remain the same. 

\begin{figure}
    \centering
    \includegraphics[width = 0.495\textwidth]{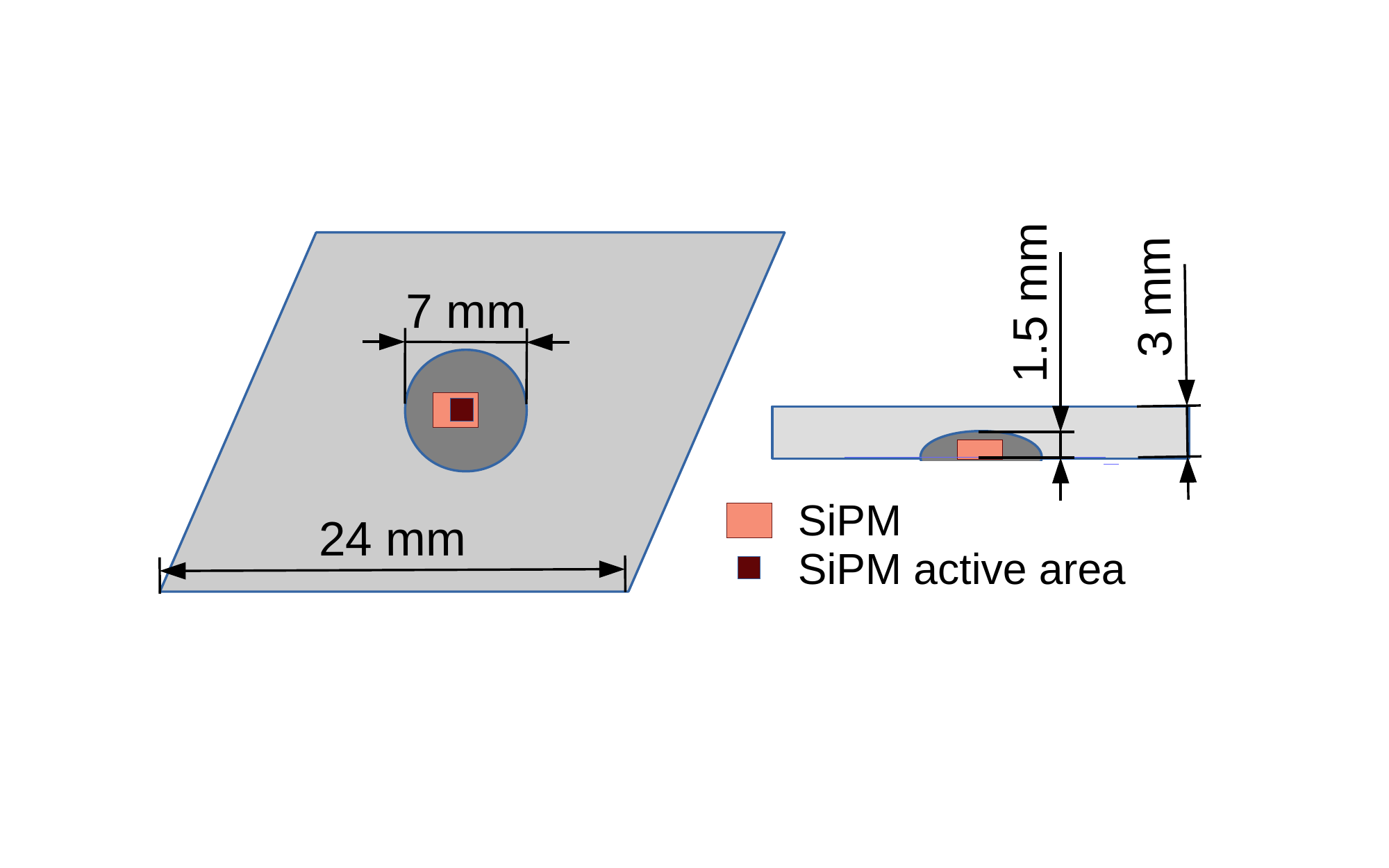}
    \hfill
    \includegraphics[width = 0.475\textwidth]{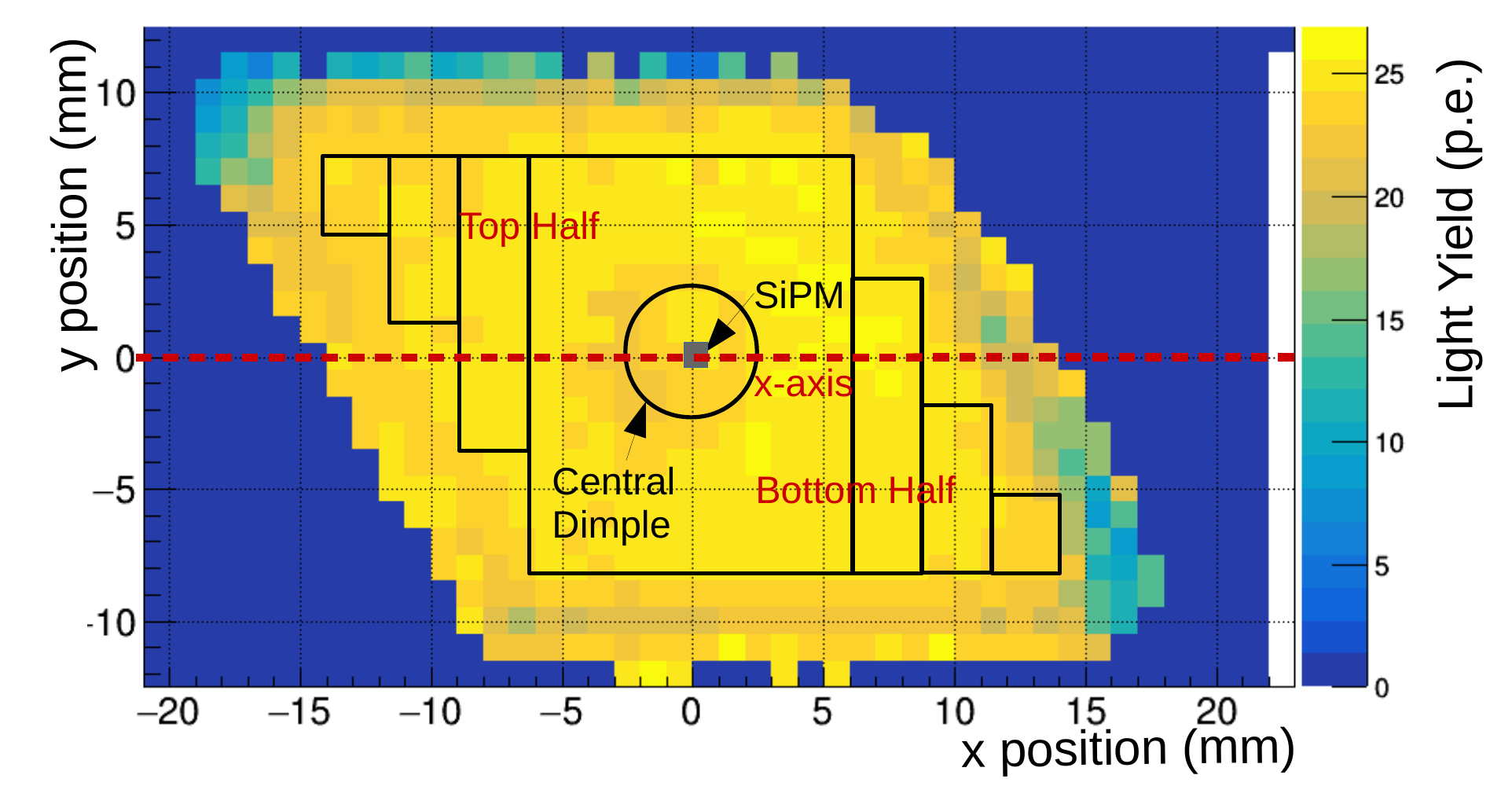}
    \caption{Illustration of the geometry including nominal SiPM placement (left) and scan results (right) for the rhomboidal tile with an edge length of 24 mm and internal angles of 60$^\circ$ and 120$^\circ$. Also the rectangles used to define the fiducial area for the calculation of average response and asymmetry are shown. The response color scale is normalized to the bin with the highest signal.}
    \label{fig:RhomTileAlignment}
\end{figure}

The same strategy is also applied to the rhomboidal tile, shown in figure \ref{fig:RhomTileAlignment}. The rhombus used here has a side length of 24 mm, and internal angles of 60$^\circ$ and 120$^\circ$. Also here, the two directions studied for the misalignment are not equivalent, since the displacements along the y axis are orthogonal to an edge, while displacements along the x axis are not.

\begin{figure}
    \centering
    \includegraphics[width = 0.495\textwidth]{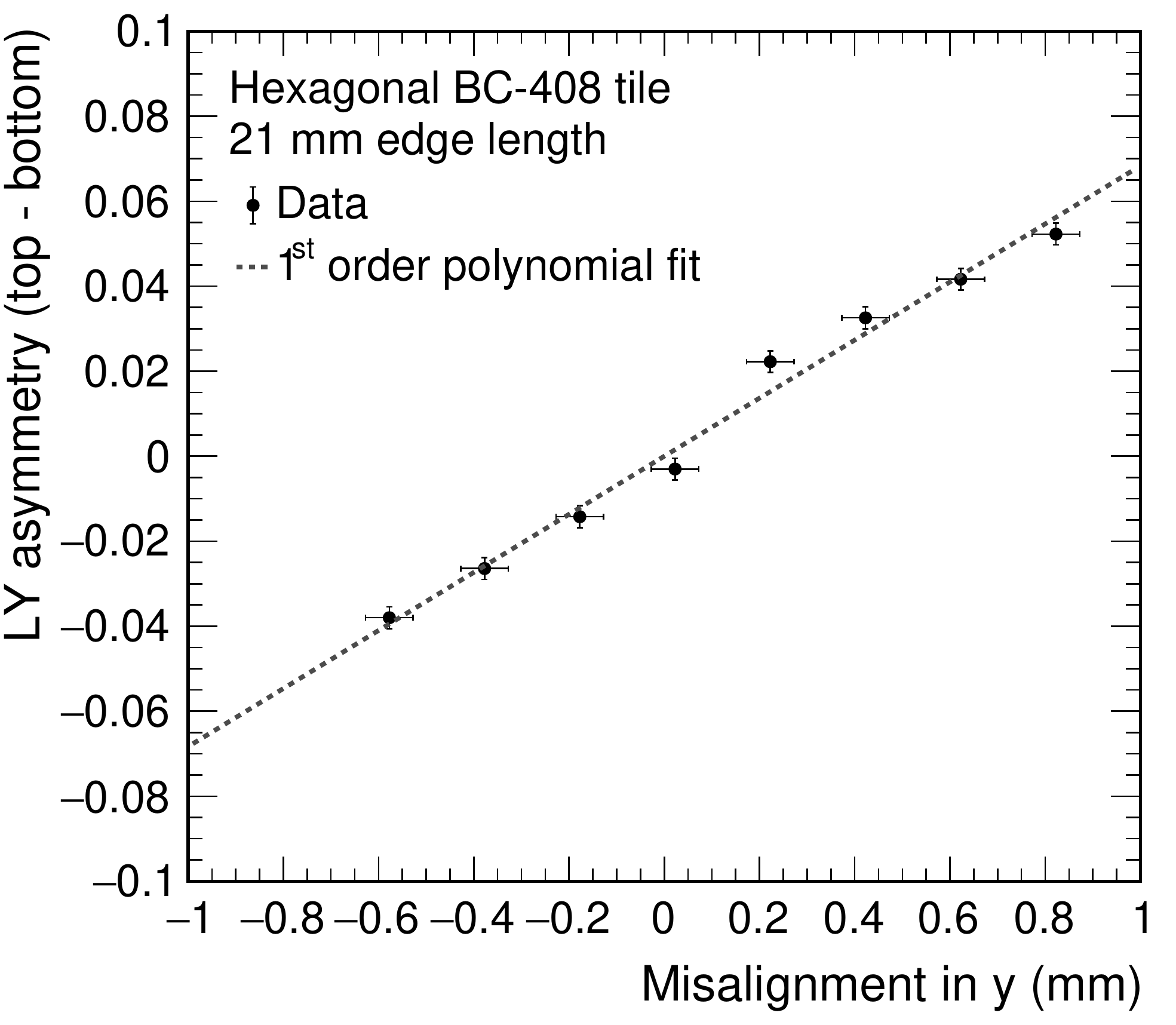}
    \hfill
    \includegraphics[width = 0.495\textwidth]{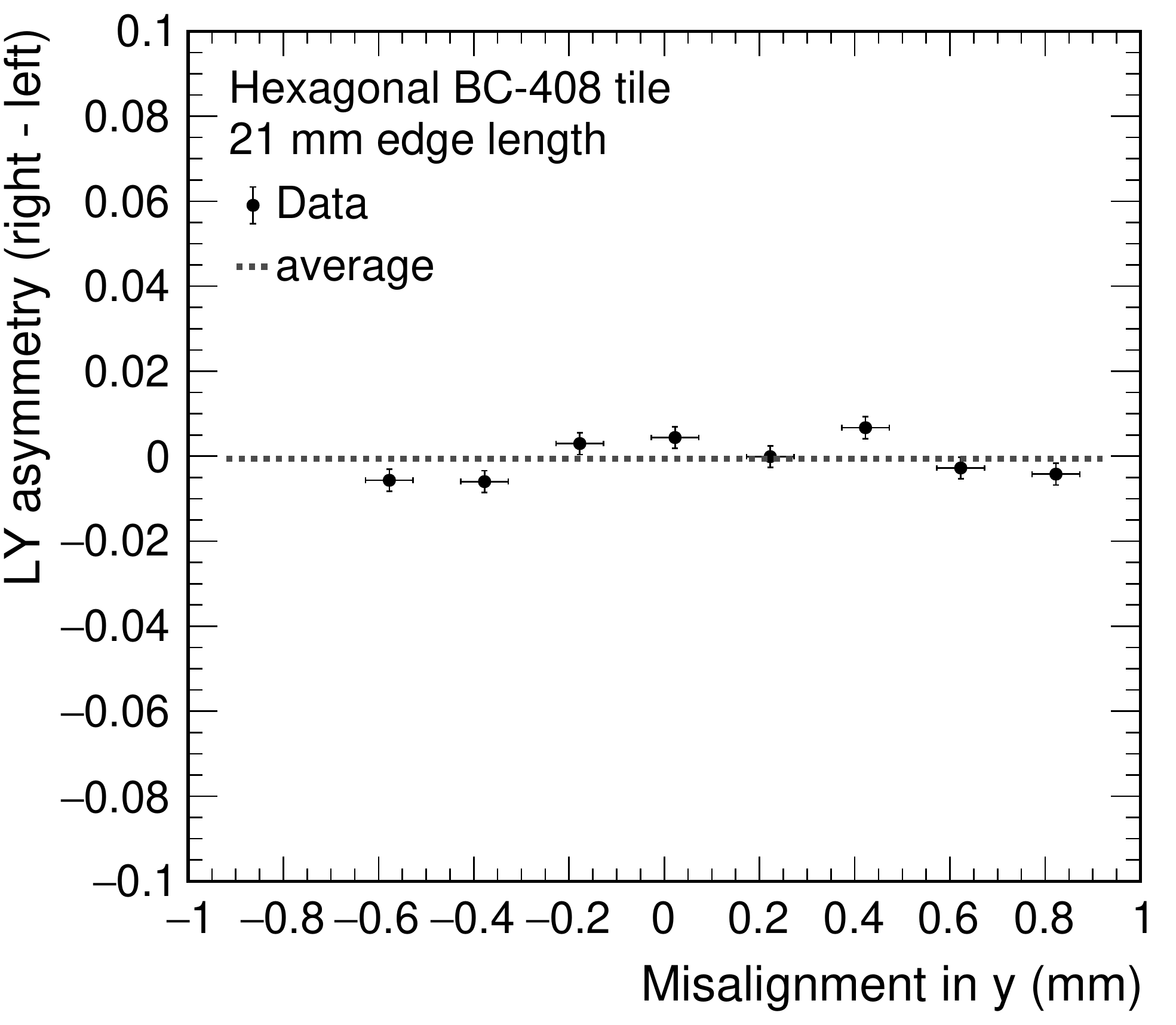}
    \caption{Measured light yield asymmetry $A_{\mathrm{hem}}$ for the 21 mm length hexagonal BC-408 tile for varying displacements in the $y$ direction. {\it Left:} Top - bottom asymmetry. The data points are fit with a first order polynomial to extract the slope as a measure for the impact of the displacement. The fitted slope is \mbox{$G_{norm}$ = 0.068 $\pm$ 0.003 mm$^{-1}$}. {\it Right:} Right - left asymmetry. The average of the data points is indicated by the horizontal line.}
    \label{fig:MeasuredAsymmetryHex}
\end{figure}

\begin{figure}
    \centering
    \includegraphics[width = 0.495\textwidth]{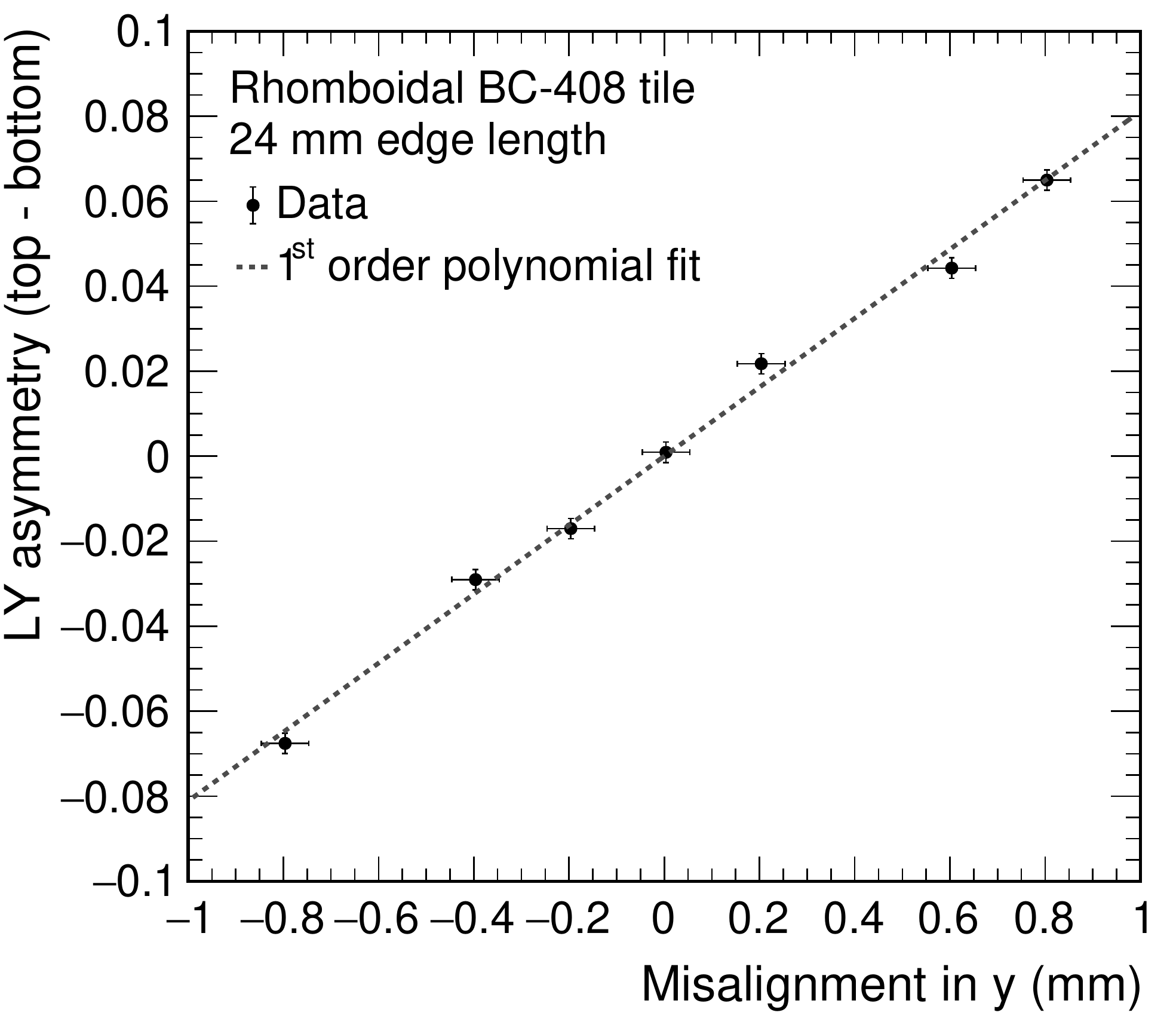}
    \hfill
    \includegraphics[width = 0.495\textwidth]{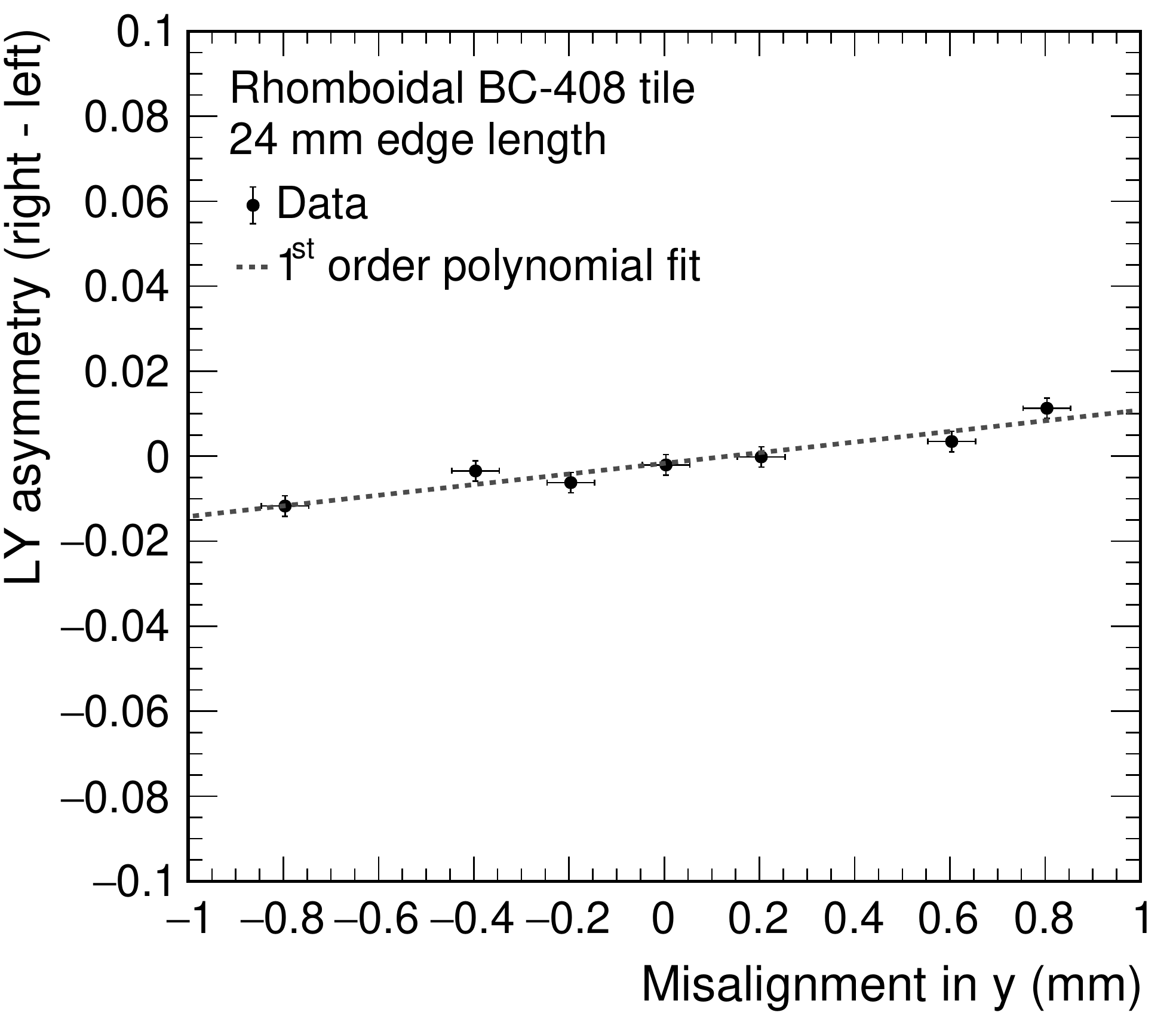}
    \caption{Measured light yield asymmetry $A_{\mathrm{hem}}$ for the 24 mm length rhomboidal BC-408 tile for varying displacements in the $y$ direction. {\it Left:} Top - bottom asymmetry. The data points are fit with a first order polynomial to extract the slope as a measure for the impact of the displacement. The fitted slope is \mbox{$G_{norm}$ = 0.081 $\pm$ 0.003 mm$^{-1}$}. {\it Right:} Right - left asymmetry, orthogonal to the axis of displacement. The data points are fit with a first order polynomial. The fitted slope is 0.013 $\pm$ 0.002 mm$^{-1}$.}
    \label{fig:MeasuredAsymmetryRh}
\end{figure}

Figure \ref{fig:MeasuredAsymmetryHex} shows the results obtained with the hexagonal tile with an edge length of 21 mm, for displacements in the $y$ direction, along the line connecting two opposite corners. As for the square tiles, a linear dependence of the asymmetry in the direction of the displacement is observed, while the asymmetry in the orthogonal direction is independent of the misalignment of photon sensor and scintillator tile. 

\begin{table}
\centering
\caption{\label{tab:ResultsHex} Summary of the results of hexagonal and rhomboidal tiles. One tile of each type has been studied. For each tile geometry, measurements are done along the two orthogonal axes. The uncertainties are the uncertainties of the fit.}
\bigskip
\begin{tabular}{|l|c|c|c|}
\hline
Tile (all BC-408) & axis & length along axis ($L_d$) [mm] & $G_{norm}$ [mm$^{-1}$] \\
\hline
\multirow{2}{*}{$14$ mm length hexagon} & x & 24.25 & 0.121 $\pm$ 0.008\\
& y & 28.0 & 0.105 $\pm$ 0.005\\
\hline
\multirow{2}{*}{$21$ mm length hexagon} & x & 36.37 & 0.077 $\pm$ 0.005\\
& y & 42.0 & 0.068 $\pm$ 0.003\\
\hline
\multirow{2}{*}{$24$ mm length rhombus} & x & 24.0 & 0.081 $\pm$ 0.003\\
& y & 20.78 & 0.078 $\pm$ 0.004\\
\hline
\end{tabular}
\end{table}

The full set of results obtained for the two hexagonal tiles is summarized in Table \ref{tab:ResultsHex}. In contrast to the observations made for square tiles, here the results do depend on the axis along which the displacement is introduced. This is expected, since the two axes considered are not equivalent, with the $x$ axis connecting the centers of two opposite edges, and the $y$ axis connecting two opposite corners. The dependence of the slope on the axis is characterized by the length of the scintillator tile along this axis, given by the distance of the two points where the axis intersects the tile sides, denoted by $L_d$. A smaller $L_d$ results in a larger slope of the asymmetry introduced by a misalignment.

The results of the displacement scan for the rhomboidal tile along the $y$ axis, which is perpendicular to two opposite edges, is shown in figure \ref{fig:MeasuredAsymmetryRh}. This tile shows a different behavior than the square and hexagonal ones, with a linear dependence of the asymmetry on the misalignment both in the direction of the displacement and orthogonal to it. The slope of the asymmetry in the orthogonal direction is significantly smaller, with approximately 20\% of the value observed in the direction of displacement. 

This difference in behavior can be understood from the geometrical properties. While square and hexagonal tiles are left-right (or top-bottom) symmetric around the axes of misalignment considered here, this is not the case for the rhombus. Thus the changes in light collection introduced by moving the photon sensor out of the center of the dimple affect the response uniformity in both directions. 

Still, also here the by far dominating effect of a misalignment is the development of an asymmetry in the direction of the displacement. As for the hexagonal tiles, the slope of this asymmetry depends on the axis, since the two axes considered are not identical, with both axes connecting the centers of opposite edges, where the $y$ axis is orthogonal to the respective edges, while the $x$ axis is not. The results obtained for the asymmetry slope in the direction of misalignment are summarized in Table \ref{tab:ResultsHex}, using $L_d$ to characterize the differences between the two axes, as for the hexagonal tiles discussed above.

\subsection{Geometry, size and material dependence of misalignment impact}

\begin{figure}[ht]
    \centering
    \includegraphics[width=0.65\textwidth]{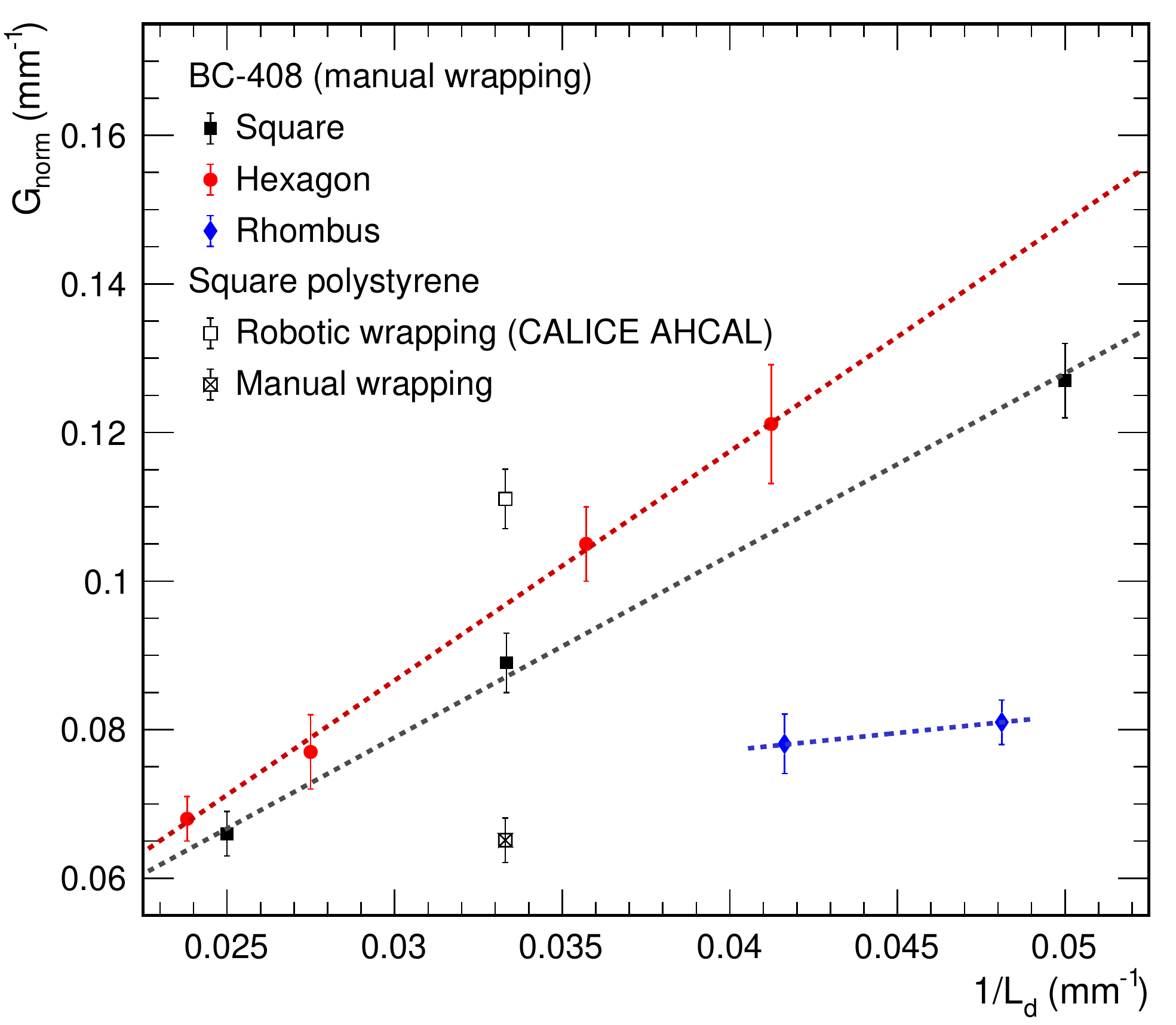}
    \caption{Gradient of the normalized asymmetry per unit length of displacement $G_{norm}$ as a function of the inverse of the length of the tile along the axis of displacement $1/L_d$ for all tile geometries, sizes and materials studied. For the types where measurements at more than one $L_d$ value exists, a fit of first order polynomial is performed to extract the slope of the dependence. The fit results are $2.45 \pm 0.23$ for the square tiles, $3.08 \pm 0.38$ for the hexagonal tiles, and $0.47 \pm 0.78$ for the rhombus. Since the two measurements of the rhombus represent a single tile, no firm conclusions can be drawn from these results.} 
    \label{fig:AsymmetryOverview}
\end{figure}

The results presented above allow to draw conclusions concerning the scaling of the effects of a misalignment with the geometry, size, and to a limited extent also with the scintillator material and reflective wrapping. In this discussion, the gradient of the light yield asymmetry per unit length of photon sensor displacement, $G_{norm}$, is taken as a measure for the sensitivity of a particular scintillator tile design to misalignment. 

The results of the square and hexagonal BC-408 tiles summarized in Tables \ref{tab:Results} and \ref{tab:ResultsHex} show a clear dependence of $G_{norm}$ on the size of the tile, with larger tiles yielding a smaller values, and thus less sensitivity to alignment inaccuracies. For the square tiles, $G_{norm}$ scales with the inverse of the square root of the area. However, the results of the hexagonal and rhomboidal show an additional dependence on the direction of the displacement, which cannot be captured as a function of the area. It is observed that the light yield asymmetry in the direction of the longer axis is smaller than the one along the shorter axis. It is thus assumed that $G_{norm}$ scales linearly with the inverse of the length of the axis $L_d$, as introduced above. 

Figure \ref{fig:AsymmetryOverview} shows all measured $G_{norm}$ values presented in the present paper versus the inverse of the respective axis length $1/L_d$. The results obtained for the square and hexagonal BC-408 tiles show very good agreement with a linear scaling, as confirmed by the fits with a first order polynomial, with a steeper dependence observed for the hexagonal tiles. Since only two data points, obtained from a single tile, exist for the rhombus, no firm conclusion concerning the accuracy of the linear scaling behavior with $1/L_d$ can be drawn for that geometry. 

The comparison of the different tile geometries presented in figure \ref{fig:AsymmetryOverview} also reveals a geometry dependence of the sensitivity to misalignment. Hexagonal tiles show an approximately 20\% larger asymmetry than square tiles, while rhomboidal tiles appear to be substantially less affected by displacements, with an approximately 30\% smaller asymmetry when considering similar-sized tiles. This suggests that rhombus-shaped tiles may be more robust against alignment inaccuracies than square or hexagonal tiles with comparable surface area. Definitive conclusions concerning rhomboidal tiles can however not been drawn, since only a single tile has been measured.     

The measurements performed with the $30\,\times\,30$ mm$^2$ square BC-408 and polystyrene tiles also provide indications of the variations of the impact of misalignment with scintillator material and reflective wrapping. As can be seen in figure \ref{fig:AsymmetryOverview}, the CALICE AHCAL polystyrene tile shows a significantly larger asymmetry when using the robotic wrapping with reflective foil compared to a manual wrapping. As discussed in section \ref{sec:SquareTiles}, a key difference between the two wrapping techniques is that the foil for the robotically wrapped tiles is closed without overlap on one of the large surfaces, with possible light leakage. Additional potential light leakage may be present at the edges due to repeated mechanical stress during the assembly and later handling. The hand-wrapped tiles, on the other hand, have overlapping foil elements, significantly reducing possible light leakage. The difference in light collection is also apparent from the differences of the average light yield observed for the two measurements, with an approximately 15\% higher light yield observed for the manual wrapping. The BC-408 tile with the same manual wrapping shows a larger susceptibility to misalignment. This suggests that scintillator properties such as material transparency, which is higher for the cast BC-408 scintillator, also have an influence on the impact of photosensor displacements from the nominal position.

\FloatBarrier
\subsection{Size and misalignment dependence of average light yields}

\begin{figure}
    \centering
    \includegraphics[width = 0.495\textwidth]{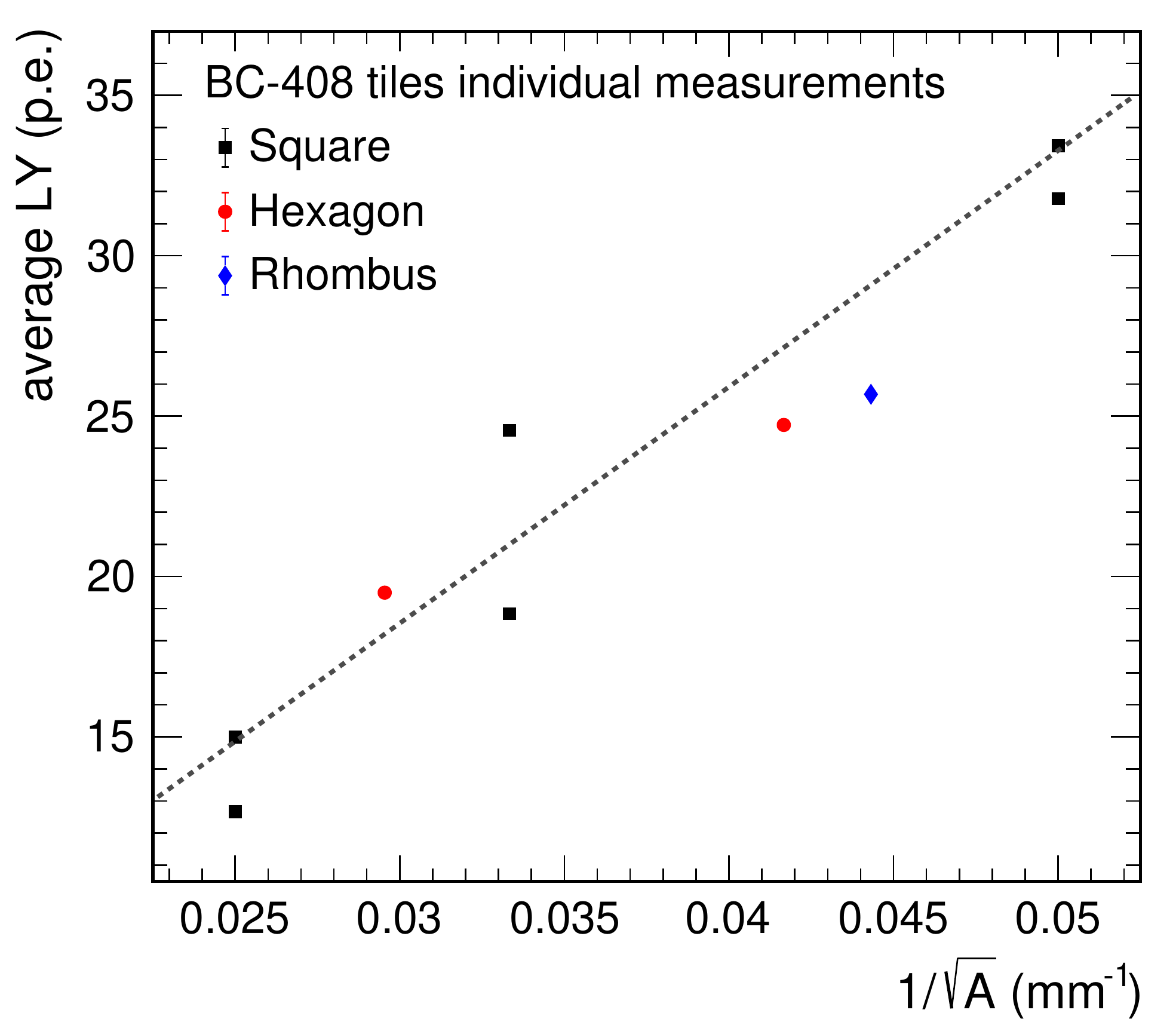}
    \hfill
    \includegraphics[width = 0.495\textwidth]{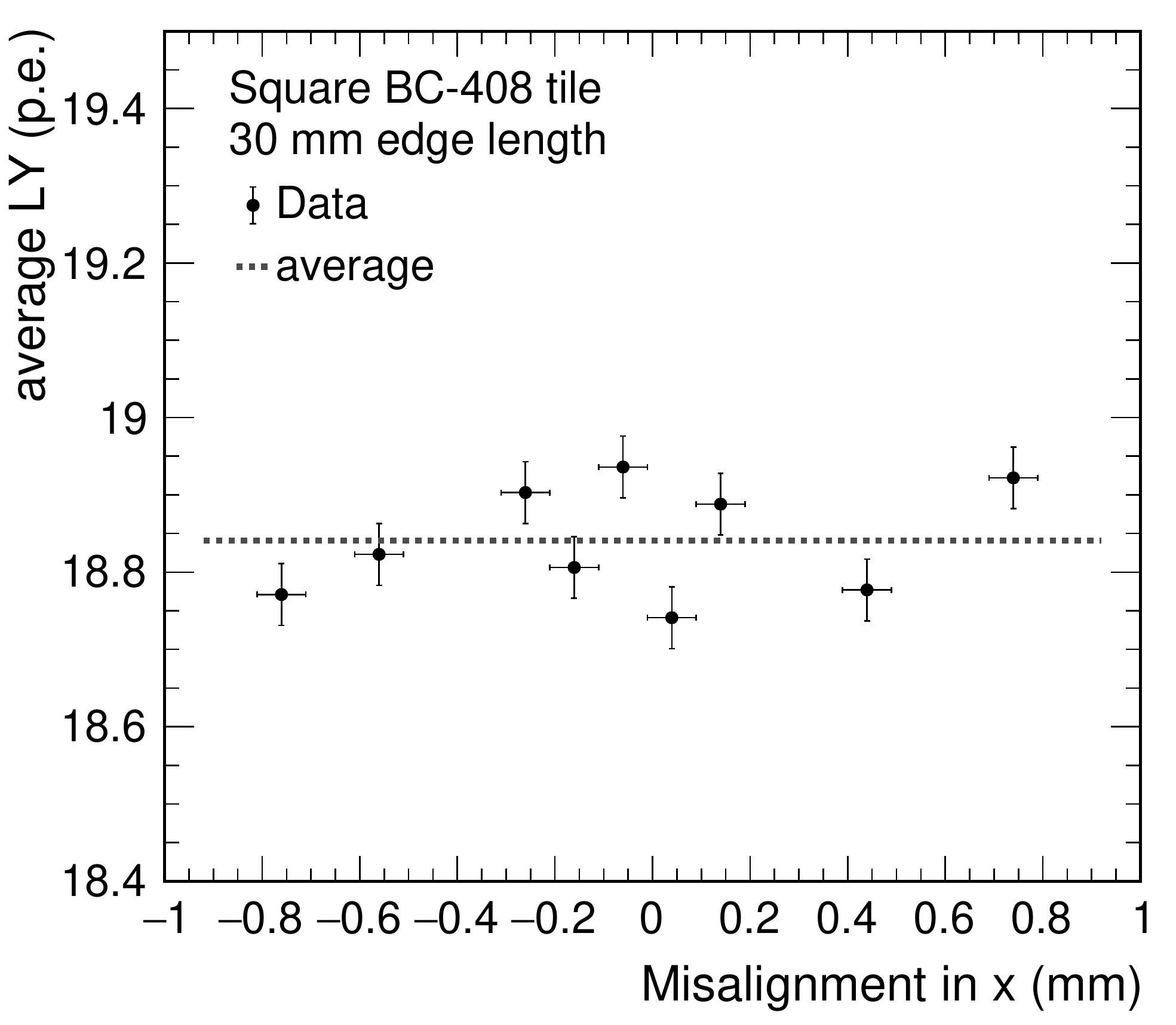}
    \caption{{\it Left:} Average light yield of all BC-408 tiles studied for the present article, shown as a function of the inverse of the square root of the tile area. The statistical uncertainties of the measurements are smaller than the marker size. Also shown is a linear fit to the data points, resulting in an offset of $-3.6\,\pm\,3.1$ p.e.\ and slope of $737\,\pm\,97$ p.e.\,mm. {\it Right:} Average light yield of one of the BC-408 square $30\,\times\,30$ mm$^2$ tiles for measurements with different misalignment along the $x$ axis. The average of all measurements is indicated by the dotted line.}
    \label{fig:AvgLYvsAreaAlignment}
\end{figure}

It has already been observed in previous studies, for example in exploratory investigations of square scintillator tiles of different sizes for a CALICE AHCAL-inspired SiPM-on-tile hadronic calorimeter for detectors at CEPC \cite{Wu:2018tiles}, that the signal amplitude depends on the tile size, with larger scintillator tiles resulting in smaller signal amplitudes. The cited publication also shows the important role that the scintillator cladding and the surface treatment play for the light yield. In the present study, the different tile sizes and geometries are used to explore the dependence of the light yield averaged over the full tile area on the size of the tile, while keeping the scintillator and tile cladding material constant. Figure \ref{fig:AvgLYvsAreaAlignment} {\it left} shows the average light yield of all BC-408 tiles measured for the present paper, as a function of the inverse of the square root of the tile area $A$. The figure shows that the light yield scales approximately linearly with $1/\sqrt{A}$, similarly to the global behaviour observed for the sensitivity to photon sensor misalignment discussed above. The scaling is approximately valid for all three geometries investigated here. The figure also shows the non-negligible variance of the light yield from tile to tile, attributed to sample-to-sample variations originating from the manual production and polishing of the scintillator material. The general scaling of the light yield with the area is extracted with a linear fit, with the fit results given in the figure caption. The poor $\chi^2$ of the fit due to the tile-to-tile variations, which are far in excess of the statistical uncertainties of the individual measurements, is taken into account in the calculation of the fit uncertainty.

The measurements of the different tiles for various degrees of misalignment have shown that the overall average response $S_{\mathrm{avg}}$ is independent of the misalignment, with an example shown for one of the BC-408 square $30\times 30$ mm$^2$ tiles for a displacement along the horizontal $x$ axis in figure \ref{fig:AvgLYvsAreaAlignment} {\it right}. This conclusion holds for all tile sizes and geometries. In practise this means that the global calibration of a SiPM-on-tile based calorimeter, which normally relies on MIP-based calibrations of the tile response, does not depend on the quality of the alignment of the tiles with respect to the photon sensor, since the relation of deposited energy to visible signal remains constant when averaging over the tile area. As a result, the only significant effect of a misalignment is the dipole asymmetry of the response over the active area of the tile, resulting in local non-uniformities that can affect the constant term of the energy resolution but leave the global response of the calorimeter unaffected. 

\section{Summary and Conclusions}

In summary, we have presented a study of the impact of a misalignment of small scintillator tiles used for SiPM-on-tile based calorimeters with respect to the photon sensor on the response uniformity. This included the investigation of square, hexagonal and rhomboidal tiles with different dimensions and scintillator material. For all geometries and sizes, a displacement results in the formation of a dipole asymmetry in the spatial response distribution, which is aligned in the direction of the displacement. The magnitude of this asymmetry scales linearly with the size of the displacement. The global response, averaged over the full active area of a tile, does not change with changing alignment, showing that displacements relative to the nominal position do not influence the calibration of larger calorimeter systems. Comparing the global light yield of different tile geometries and sizes shows that this quantity scales approximately with the inverse of the square root of the area of the tile for all studied geometries, with smaller scintillator tiles resulting in larger signals. In view of the non-negligible tile-to-tile variations observed, a more accurate determination of the scaling properties will require the investigation of larger samples. 

The size of the observed dipole asymmetry for a unit displacement depends on the length of the tile along the axis of movement, scaling with the inverse of the axis length. Smaller scintillator tiles are thus more susceptible to misalignment than larger tiles. 
For scintillator tiles with a size of around 9 cm$^2$, as considered for hadron calorimeters at future linear electron-positron colliders, the typical dipole asymmetry introduced by a displacement of the tile with respect to the photon sensor is around 0.1 per 1 mm of displacement. This means that for such a displacement, the average signal amplitude of one half of the tile is 10\% higher than that of the other half. Considering that approximately 80\% of the active area of typical SiPM-on-tile designs show a response within 5\% of the mean  \cite{Liu:2015cpe}, limiting the displacement-induced asymmetry to below 0.05 would exclude significant light yield distortions beyond those intrinsic in the scintillator tile design. This translates to a limit of approximately 500 $\upmu$m on the tile misalignment, a value compatible with achievable tile production and board assembly tolerances in mass production techniques. 

The measurements presented here also indicate a scintillator-material and cladding dependence of the tile properties studied in the present paper, which would be an interesting subject for further systematic investigation.

\acknowledgments

We thank our colleagues in the CALICE Collaboration for helpful discussions. This project has received funding from the European Union's Horizon 2020 Research and Innovation programme under Grant Agreement no.\ 654168.

\bibliography{TileTolerance}

\providecommand{\href}[2]{#2}\begingroup\raggedright\begin{thebibliography}{10}

\bibitem{Adloff:2010hb}
{\scshape CALICE} collaboration, C.~Adloff et~al., \emph{{Construction and
  Commissioning of the CALICE Analog Hadron Calorimeter Prototype}},
  \href{http://dx.doi.org/10.1088/1748-0221/5/05/P05004}{\emph{JINST}
  {\bfseries 5} (2010) P05004},
  [\href{https://arxiv.org/abs/1003.2662}{{\ttfamily arXiv:1003.2662}}].

\bibitem{Sefkow:2018rhp}
{\scshape CALICE} collaboration, F.~Sefkow and F.~Simon, \emph{{A highly
  granular SiPM-on-tile calorimeter prototype}},
  \href{http://dx.doi.org/10.1088/1742-6596/1162/1/012012}{\emph{J. Phys. Conf.
  Ser.} {\bfseries 1162} (2019) 012012},
  [\href{https://arxiv.org/abs/1808.09281}{{\ttfamily arXiv:1808.09281}}].

\bibitem{Collaboration:2293646}
{CMS Collaboration}, \emph{{The Phase-2 Upgrade of the CMS Endcap
  Calorimeter}}, {\emph{{\href{https://cds.cern.ch/record/2293646}{\normalfont
  CERN-LHCC-2017-023}}} (2017) }.

\bibitem{Emberger:2018pgr}
L.~Emberger and F.~Simon, \emph{{A highly granular calorimeter concept for long
  baseline near detectors}},
  \href{http://dx.doi.org/10.1088/1742-6596/1162/1/012033}{\emph{J. Phys. Conf.
  Ser.} {\bfseries 1162} (2019) 012033},
  [\href{https://arxiv.org/abs/1810.03677}{{\ttfamily arXiv:1810.03677}}].

\bibitem{Blazey:2009zz}
G.~Blazey et~al., \emph{{Directly Coupled Tiles as Elements of a Scintillator
  Calorimeter with MPPC Readout}},
  \href{http://dx.doi.org/10.1016/j.nima.2009.03.253}{\emph{Nucl. Instrum.
  Meth.} {\bfseries A605} (2009) 277--281}.

\bibitem{Liu:2015cpe}
Y.~Liu et~al., \emph{{A Design of Scintillator Tiles Read Out by
  Surface-Mounted SiPMs for a Future Hadron Calorimeter}},  in \emph{{2014 IEEE
  Nuclear Science Symposium and Medical Imaging Conference and 21st Symposium
  on Room-Temperature Semiconductor X-ray and Gamma-ray Detectors (NSS/MIC 2014
  / RTSD 2014) Seattle, WA, USA, November 8-15, 2014}},
  \href{https://arxiv.org/abs/1512.05900}{{\ttfamily arXiv:1512.05900}}.

\bibitem{Simon:2010hf}
F.~Simon and C.~Soldner, \emph{{Uniformity Studies of Scintillator Tiles
  directly coupled to SiPMs for Imaging Calorimetry}},
  \href{http://dx.doi.org/10.1016/j.nima.2010.03.142}{\emph{Nucl.Instrum.Meth.}
  {\bfseries A620} (2010) 196--201},
  [\href{https://arxiv.org/abs/1001.4665}{{\ttfamily arXiv:1001.4665}}].

\bibitem{Munwes:2634923}
Y.~Munwes, P.~Chau and F.~Simon, \emph{{Performance of test infrastructure for
  highly granular optical readout}},
  {\emph{{\href{http://cds.cern.ch/record/2634923}{\normalfont
  AIDA-2020-D14.2}}} (2018) }.

\bibitem{Chadeeva:2018ezg}
M.~Chadeeva, S.~Korpachev, V.~Rusinov and E.~Tarkovskii, \emph{{Tests of
  Scintillator Tiles for the Technological Prototype of Highly Granular Hadron
  Calorimeter}}, \href{http://dx.doi.org/10.18502/ken.v3i1.1768}{\emph{KnE
  Energ.\ Phys.} {\bfseries 3} (2018) 363--371}.

\bibitem{Wu:2018tiles}
Z.~Wu et~al., \emph{{Studies of detector cells for a hadronic calorimeter based
  on plastic scintillators}},
  \href{http://dx.doi.org/10.1007/s41605-018-0051-5}{\emph{Radiat.\ Detect.\
  Technol.\ Methods} {\bfseries 2} (2018) 22}.

\end{thebibliography}\endgroup

\end{document}